\documentclass{aa}
\usepackage{graphicx}
\usepackage{txfonts}
\begin{document}
   \title{Dust properties of UV bright galaxies at $z \sim 2$
   \thanks{Based on observations obtained with FORS at the VLT, Paranal, Chile
           on the course of the observing proposals 63.O-0005, 64.O-0149, 
           64.O-0158, 65.O-0049, 66.A-0547, 68.A-0013, 68.A-0014, 69.A-0104}}

   \author{S. Noll\inst{1}
   \and D. Pierini\inst{1}}

   \offprints{S. Noll}

   \institute{Max-Planck-Institut f\"ur extraterrestrische Physik, 
              Giessenbachstr., 85748 Garching, Germany\\
              \email{snoll@mpe.mpg.de}
             }

   \date{Received; accepted}

   \abstract{
We investigate the properties of the extinction curve in the rest-frame UV 
for a sample of 34 UV-luminous galaxies at $2 < z < 2.5$, selected from the 
FORS Deep Field (FDF) spectroscopic survey. A new parametric description of 
the rest-frame UV spectral energy distribution is adopted; its sensitivity to 
properties of the stellar populations or of dust attenuation is established 
with the use of models. The latter are computed by combining composite stellar 
population models and calculations of radiative transfer of the stellar and 
scattered radiation through the dusty interstellar medium (ISM) for a 
dust/stars configuration describing dust attenuation in local starbursts. In 
the favoured configuration the stars are enveloped by a shell with a 
two-phase, clumpy, dusty ISM. The distribution of the $z \sim 2$ UV-luminous 
FDF galaxies in several diagnostic diagrams shows that their extinction 
curves range between those typical of the Small and Large Magellanic Clouds 
(SMC and LMC, respectively). For the majority of strongly reddened objects 
having a UV continuum slope $\beta > -0.4$ a significant 2175\,\AA{} 
absorption feature (or ``UV bump'') is inferred, indicating an LMC-like 
extinction curve. On the other hand, the UV continua of the least reddened 
objects are mostly consistent with SMC-like extinction curves, lacking a 
significant UV bump, as for the sample of local starbursts investigated by 
Calzetti and collaborators. Furthermore, the most opaque ($\beta \sim 0$) 
and, thus (for our models), dustiest UV-luminous FDF galaxies tend to be 
among the most metal rich, most massive, and largest systems at $z \sim 2$, 
indicating $\langle Z \rangle \sim 0.5 - 1$\,Z$_{\odot}$, 
$\langle M_{\rm stars} \rangle \sim 6 \times 10^{10}$\,M$_{\odot}$, and
$\langle R_{\rm eff} \rangle \sim 4$\,kpc, respectively. The presence of the 
UV bump does not seem to depend on the total metallicity, as given by the 
equivalent width (EW) of the C\,IV doublet. Conversely, it seems to be 
associated with a large average EW of the six most prominent interstellar
low-ionisation absorption lines falling in the FORS spectra. The average EW 
of these saturated lines offers a proxy for the ISM topology. We interpret 
these results as the evidence for a difference in the properties of the dusty 
ISM among the most evolved UV-luminous, massive galaxies at $z \sim 2$.

   }

   \maketitle
%
%________________________________________________________________

\section{Introduction}\label{introduction}

The electromagnetic radiation propagating through a medium containing dust 
particles experiences extinction. In general, the intensity of the 
transmitted beam is reduced by two physical processes: absorption and 
scattering. Dust grains absorb and scatter more the shorter the wavelength.
As a consequence, both the total intensity and the spectrum of the pristine 
radiation are modified. The so-called {\it extinction curve} quantifies the 
amount of extinction as a function of wavelength along the line of sight.
It depends on the chemical composition, structure, and size distribution
of the intervening dust grains (Mathis et al. \cite{MAT77}; D\'esert et al. 
\cite{DES90}; Dweck et al. \cite{DWE97}; Draine \& Li \cite{DRA01}; Li \& 
Draine \cite{LI01}; Zubko et al. \cite{ZUB04}). Hence, studies of extinction 
provide information on the properties of dust.

The {\it direct} determination of the spectral dependence of extinction from 
observational data is limited to our own Galaxy, the Magellanic Clouds, and 
M\,31 (for reviews see Fitzpatrick \cite{FIT04} and Clayton \cite{CLA04}).

Extinction curves of our own Galaxy vary for different sight lines and 
towards different environments (from dense molecular clouds to the diffuse 
interstellar medium, ISM; see Fitzpatrick \cite{FIT04}). However, 
the extinction curves determined for the diffuse ISM of the Milky Way (MW) 
seem to represent a one-parameter family, the parameter being the 
total-to-selective extinction ratio $R_V = A_V/E_{B-V}$ (Cardelli et al. 1989, 
\cite{CAR89}; Fitzpatrick \cite{FIT99}; Clayton et al. \cite{CLA00}; 
Valencic et al. \cite{VAL04}; see however Fitzpatrick \cite{FIT04}). Broadly 
speaking, $R_V$ is a proxy for the average grain size: low-$R_V$ sight lines 
have more small grains than high-$R_V$ sight lines. The smaller the grains, 
the steeper the extinction curve in the ultraviolet (UV).

The average extinction curve for the diffuse ISM of the MW has $R_V = 3.1$.
It is characterised by the presence of a bump at a constant central 
wavelength of 2175\,\AA{}. This absorption feature (Witt \& Lillie 
\cite{WIT73}) is most likely produced by graphitic carbon, though its origin 
is still under discussion (Draine \cite{DRA03}; Whittet \cite{WHI03} and 
references therein). The width of the UV bump seems to depend on the 
environment: narrow bumps tend to be associated with diffuse regions, while 
broad bumps favour dense regions (Valencic et al. \cite{VAL04}). The 
occurrence of weak bumps may reflect processing that modifies or destroys the 
bump carrier in dark clouds (Whittet et al. \cite{WHI04}).

The few sight lines studied in M\,31 seem to show a \cite{CAR89} far-UV 
extinction and a weak 2175\,\AA{} bump (Bianchi et al. \cite{BIA96}). There 
are remarkable differences in dust grain properties between the Milky Way and 
the Large and Small Magellanic Clouds (LMC and SMC, respectively). The sight 
lines toward the LMC\,2 supershell and the average LMC extinction curve 
exhibit a weaker bump and stronger far-UV extinction than the average MW 
$R_V = 3.1$ extinction curve. The sight lines toward the SMC bar exhibit 
extremely weak or absent UV-bump absorption and a very strong far-UV 
extinction, which is indicative of a very large fraction of small dust 
particles (Rodrigues et al. \cite{ROD97}). These differences may reflect 
differences in metallicity\footnote{The LMC and SMC are deficient in heavy 
elements by factors of about $2.5$ and 7, respectively, compared with the 
solar standard (Westerlund \cite{WES97}).} and gas-to-dust ratio (Bohlin et 
al. \cite{BOH78}; Luck \& Lambert \cite{LUC92}; Gordon et al. \cite{GOR03}).

Different environments, such as star-formation regions where large amounts of 
UV radiation and shocks are present, may play a fundamental role in 
processing dust. This is not easy to establish however. As remarked by 
Clayton (\cite{CLA04}), the LMC\,2 supershell is located in the 30\,Dor 
region, which is a much larger star-forming region than any in the SMC, but 
the LMC\,2 supershell extinction curve is in much better agreement with a 
\cite{CAR89} MW extinction curve than the SMC extinction curve. Nevertheless, 
it is intriguing that nearby starburst galaxies spanning a large range in 
metallicity (from sub- to super-solar) possibly contain dust with an 
extinction curve lacking a 2175\,\AA{} bump, like the SMC curve, and a 
steep far-UV rise, intermediate between the MW and SMC curves (Calzetti et al. 
\cite{CAL94}, \cite{CAL00}; Gordon et al. \cite{GOR97}; Leitherer et al. 
\cite{LEI02}). For these objects, the properties of dust (i.e. the extinction 
curve) can be determined only in an {\it indirect} and rough way, after 
resorting to models of radiative transfer of the stellar and scattered 
radiation through a dusty ISM (Witt \& Gordon 2000, \cite{WIT00}; Misselt et 
al. \cite{MIS01}; Pierini et al. \cite{PIE04}; Inoue \cite{INO05}). In fact, 
on a galaxy scale, the propagation of electromagnetic radiation through a 
dusty medium is described by the {\it attenuation function}, i.e. the 
combination of the extinction curve with the geometry of the system, in which 
a substantial fraction of the scattered light is returned to the observer's 
line of sight.

SMC-like dust seems to characterise also starburst galaxies in the 
distant universe. This holds for individual objects with a well-sampled SED 
(e.g. Gordon et al. \cite{GOR99}) as well as for large photometric samples, 
like that of $\sim 1000$ Lyman-break galaxies (LBGs) at $z \sim 3$ observed 
by Steidel et al. (\cite{STE03}) (Vijh et al. \cite{VIJ03}). The absence of a 
2175\,\AA{} bump seems to characterise the rest-frame UV spectrum of other 
objects lying at intermediate/high redshift, like quasars/AGN (Pitman et al. 
\cite{PIT00}; Maiolino et al. \cite{MAI01}; Hopkins et al. \cite{HOP04}) and 
GRB-host galaxies (e.g. GRB\,020813 at $z = 1.255$, Savaglio \& Fall 
\cite{SAV04}). On the other hand, a UV bump seems to be detected in the 
composite absorption spectrum of 96 intervening Mg\,II systems at 
$0.2 < z < 2.2$ (Malhotra \cite{MAL97}). However, Khare et al. (\cite{KHA05}) 
come to an opposite conclusion for the composite absorption spectrum of 815 
intervening QSO absorbers at $1 \la z \la 1.9$. Further evidence for MW-type 
dust being also present at high redshift comes from a composite spectrum of 
nine radio galaxies at $z \sim 2.5$ (Vernet et al. \cite{VER01}), and from the 
individual absorption spectra of three intervening Mg\,II absorbers at 
$1.4 \la z \la 1.5$ (Wang et al. \cite{WAN04}).

It is not clear when star-forming galaxies at high redshift exhibit an 
extinction curve with a 2175\,\AA{} bump. Therefore, we have selected a 
sample of 34 UV-luminous galaxies at $2 < z < 2.5$ from the FORS Deep Field 
spectroscopic survey (Noll et al. \cite{NOL04}; see Sect.~\ref{data}). We 
will characterise their composite extinction curve at rest-frame UV 
wavelengths in terms of far-UV slope and presence/absence of the UV bump. The 
UV-luminous galaxies at $z \sim 2$ exhibit large values of UV continuum 
reddening (e.g. van Dokkum et al. \cite{DOK03}; Daddi et al. \cite{DAD04}; 
Noll et al. \cite{NOL04}), if the slopes of their rest-frame UV continuum 
spectra are interpreted as for local starburst galaxies (Meurer et al. 
\cite{MEU99}; Calzetti et al. \cite{CAL00}; Leitherer et al. \cite{LEI02}). 
In this case, one can obtain attenuation factors at 1500\,\AA{} up to 
$\sim 10^2$ (Noll et al. \cite{NOL04}), that are more than one order of 
magnitude larger than those determined for nearby starbursts, and correspond 
to the largest values estimated for LBGs at $z \sim 3$ by Vijh et al. 
(\cite{VIJ03}). At variance with those LBGs at $z \sim 3$, the UV-luminous 
galaxies at $z \sim 2$ are characterised by half-light radii of several kpc, 
total stellar masses between a few $10^{10}$ and more than 
$10^{11}$\,M$_{\odot}$, metallicity between the LMC one and solar, and the 
absence of significant Ly$\alpha$ emission (Mehlert et al. \cite{MEH02}; Daddi 
et al. \cite{DAD04}; Noll et al. \cite{NOL04}; Shapley et al. \cite{SHA04}; 
Steidel et al. \cite{STE04}; van Dokkum et al. \cite{DOK04}; Drory et al. 
\cite{DRO05}). Hence, the properties of the mixture of dust grains present in 
UV-luminous galaxies at $z \sim 2$ may differ from those of LBGs at $z \sim 3$ 
and those of local starburst galaxies.

In the following we illustrate this by making use of a new parametric 
description of the rest-frame UV SED that allows us to define dust-sensitive 
diagnostic diagrams. We interpret the distribution of the FDF sample galaxies 
in each of these diagrams through a comparison with that obtained either for 
a local reference sample of starburst galaxies or for models of starburst 
galaxies combining stellar population synthesis and radiative transfer 
calculations (Sect.~\ref{analysis}). 

Throughout this paper $H_0 = 70\,{\rm km}\,{\rm s}^{-1}\,{\rm Mpc}^{-1}$, 
$\Omega_{\Lambda} = 0.7$, and $\Omega_{\rm M} = 0.3$ are adopted.

\section{The spectroscopic sample: selection and data}\label{data}

\subsection{The FORS sample of UV-luminous galaxies at $z \sim 2$}\label{FDF}

\begin{table}
\caption[]{The spectroscopic sample of FDF galaxies at $2 < z < 2.5$ (Noll et 
al. \cite{NOL04}). IDs, world coordinates, and total $I$ magnitudes were 
taken from Heidt et al. (\cite{HEI03}). Redshifts were derived by Noll et al. 
(\cite{NOL04}). Luminosities at 1500\,\AA{} were taken from the same source,
though corrected for the modified cosmology used.}
\label{tab_FDF_basic}
\centering
\begin{tabular}{c c c c c c}
\hline\hline
\noalign{\smallskip}
No. & RA & DEC & $I_{\rm Vega}$ & $z$ & $\log L_{1500}$ \\
(FDF) & (J2000) & (J2000) & [mag] & & [W/\AA{}] \\
\noalign{\smallskip}
\hline
\noalign{\smallskip}
1208 & 01 05 51.9 & -25 48 04 & 23.68 & 2.178 & 33.88 \\
1691 & 01 05 53.7 & -25 45 29 & 23.89 & 2.344 & 33.94 \\
1744 & 01 05 53.9 & -25 46 06 & 24.10 & 2.374 & 33.91 \\
1991 & 01 05 54.8 & -25 46 16 & 24.33 & 2.334 & 33.85 \\
2274 & 01 05 55.9 & -25 44 34 & 23.34 & 2.253 & 34.18 \\
2418 & 01 05 56.4 & -25 45 12 & 23.16 & 2.332 & 34.25 \\
2495 & 01 05 56.7 & -25 43 44 & 23.31 & 2.453 & 34.25 \\
2636 & 01 05 57.3 & -25 44 02 & 23.43 & 2.253 & 34.01 \\
3005 & 01 05 58.6 & -25 48 14 & 23.51 & 2.253 & 34.03 \\
3163 & 01 05 59.2 & -25 45 38 & 23.35 & 2.441 & 34.21 \\
3300 & 01 05 59.6 & -25 46 30 & 23.91 & 2.375 & 34.07 \\
3374 & 01 05 59.9 & -25 45 11 & 23.34 & 2.386 & 34.27 \\
3688 & 01 06 00.9 & -25 47 05 & 24.07 & 2.375 & 34.00 \\
3810 & 01 06 01.3 & -25 45 28 & 22.67 & 2.372 & 34.44 \\
3874 & 01 06 01.5 & -25 45 46 & 23.30 & 2.483 & 34.23 \\
3875 & 01 06 01.5 & -25 47 34 & 24.53 & 2.243 & 33.84 \\
3958 & 01 06 01.8 & -25 44 29 & 23.87 & 2.130 & 33.97 \\
4795 & 01 06 04.8 & -25 47 14 & 23.31 & 2.159 & 34.13 \\
4871 & 01 06 05.1 & -25 46 04 & 23.39 & 2.472 & 34.26 \\
4996 & 01 06 05.5 & -25 46 28 & 23.25 & 2.028 & 34.04 \\
5058 & 01 06 05.7 & -25 46 26 & 23.34 & 2.027 & 34.15 \\
5135 & 01 06 06.0 & -25 44 44 & 23.62 & 2.346 & 34.02 \\
5165 & 01 06 06.1 & -25 44 43 & 23.26 & 2.346 & 34.34 \\
5190 & 01 06 06.1 & -25 44 43 & 24.39 & 2.347 & 33.83 \\
5227 & 01 06 06.3 & -25 43 52 & 23.85 & 2.399 & 34.08 \\
6024 & 01 06 09.2 & -25 48 14 & 22.00 & 2.372 & 34.68 \\
6372 & 01 06 10.5 & -25 48 29 & 23.38 & 2.349 & 34.21 \\
6407 & 01 06 10.6 & -25 45 32 & 23.59 & 2.162 & 33.95 \\
6934 & 01 06 12.4 & -25 44 57 & 22.90 & 2.445 & 34.43 \\
6947 & 01 06 12.4 & -25 48 15 & 23.83 & 2.357 & 33.97 \\
7029 & 01 06 12.7 & -25 45 58 & 23.63 & 2.374 & 34.12 \\
7078 & 01 06 12.8 & -25 46 01 & 23.97 & 2.378 & 34.00 \\
7307 & 01 06 13.6 & -25 47 25 & 24.07 & 2.438 & 34.01 \\
7342 & 01 06 13.7 & -25 46 13 & 23.80 & 2.375 & 34.06 \\
\noalign{\smallskip}
\hline
\end{tabular}
\end{table}

\begin{figure}
\centering 
\includegraphics[width=8.8cm,clip=true]{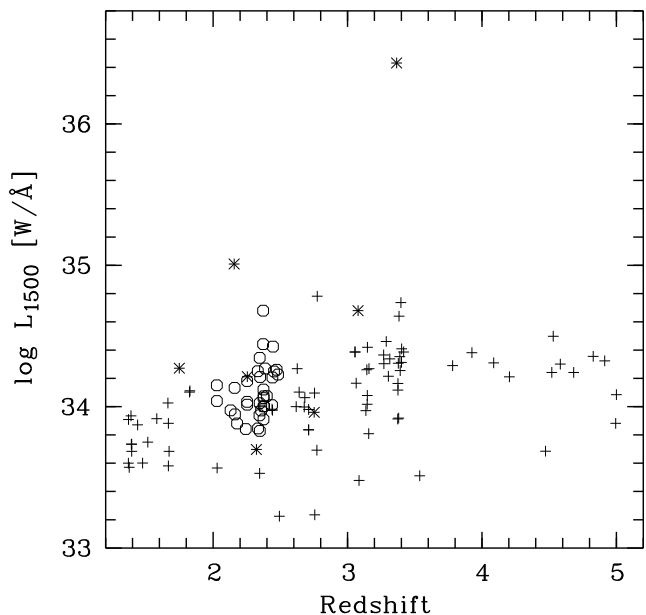}
\caption[]{The luminosity at 1500\,\AA{} (rest frame) $L_{1500}$ as a 
function of redshift for the 118 galaxies at $z > 1.35$ in the FDF 
spectroscopic sample (from Noll et al. \cite{NOL04}). The 34 galaxies under 
study (see Table~\ref{tab_FDF_basic}) are reproduced with circles. Among the 
remaining high-redshift objects, star-forming galaxies are marked by crosses 
and QSOs by asterisks. $L_{1500}$ refers to the measured emission within the 
rest-frame wavelength range from 1480 to 1520\,\AA{}.}
\label{fig_luvz}
\end{figure}

We have selected 34 galaxies (see Table~\ref{tab_FDF_basic}) with 
$I_{\rm Vega} \lesssim 24.5$ in the spectroscopic-redshift range 
$2 < z < 2.5$ from the catalogue of the FORS Deep Field (FDF) spectroscopic 
survey (Noll et al. \cite{NOL04}). The redshift interval chosen guarantees 
the covering of the rest-frame UV wavelength range from Ly$\alpha$ to around 
2500\,\AA{}, including the 2175\,\AA{} feature, with the available optical 
spectroscopy. The limiting $I$-band magnitude prevents the contamination from 
faint Ly$\alpha$ emitters that were discovered serendipitously. With an 
average luminosity at 1500\,\AA{} (rest frame) 
$\langle L_{1500} \rangle = 1.4 \times 10^{34}$\,W/\AA{} (not corrected for 
dust attenuation), the present sample of 34 galaxies represents the bright 
end of the distribution in $L_{1500}$ of FDF galaxies at $z \sim 2$ (see 
Fig.~\ref{fig_luvz}). Hereafter it is referred to as the FDF sample of 
``$z \sim 2$ UV-luminous galaxies''.

The VLT-FORS spectra of our sample galaxies were obtained with the grism 
150\,I and a $1''$ slit width (Noll et al. \cite{NOL04}), resulting in a 
measured spectral resolution element between 18 and 23\,\AA{} that
corresponds to rest-frame 5 to 8\,\AA{}. The spectra cover the entire optical 
wavelength range providing reasonable data between 3500 and 9300\,\AA{} 
(observed frame), which allows us to probe the UV SED between 1216\,\AA{} 
(i.e. the Ly$\alpha$ transition of hydrogen) and $\sim 2500$\,\AA{}, 
bracketing the region of the UV bump. Typical exposure times of about 10\,hr 
led to an average signal-to-noise ratio S/N of about 13 for the programme 
galaxies (see Noll et al. \cite{NOL04}). The quality of the spectra is 
sufficiently high for the purpose of this study. The spectra used are S/N 
optimised co-added, flux calibrated (including a correction for slit losses), 
and corrected for Galactic extinction ($E_{B-V} = 0.018$, Schlegel et al. 
\cite{SCHL98}) using the \cite{CAR89} MW extinction law. Finally, the 
atmospheric B and A absorption bands were removed from the observed spectra 
using an empirical band model. The final spectra are slightly contaminated by 
a second order spectrum for $\lambda > 6500$\,\AA{}. This effect does not 
significantly influence our study of the rest-frame UV continuum; the 
resulting relative flux uncertainties are not greater than 2 to 3\% in the 
UV-bump region for the selected galaxies.

\subsection{The IUE sample of nearby starbursts}\label{IUE}

\begin{table}
\caption[]{The comparison sample of local starburst galaxies having IUE 
spectra with suitable S/N and spectral coverage. The (activity) class 
parameter shown is taken from Kinney et al. (\cite{KIN93}) with the exception
of IRAS\,08339+6517 and NGC\,1741, for which this parameter comes from
Leitherer et al. (\cite{LEI02}) and Johnson et al. (\cite{JOH99}),
respectively. The meaning of the different parameters is as follows: 
BC(D)G~= blue compact (dwarf) galaxy; Sy\,2~= Seyfert\,2 galaxy; SB\,nuc.~= 
galaxy experiencing a starburst in its nucleus; Hs~= hot spot galaxy; Lin~= 
LINER; H\,II~= galaxy with a spectrum typical of a H\,II region. All 
redshifts are taken from the NASA/IPAC Extragalactic Database (NED). Values 
of $E_{B-V}$ were calculated using the maps of Galactic dust IR emission of 
Schlegel et al. (\cite{SCHL98}). The oxygen abundances $O/H = 12 + \log(O/H)$ 
are taken from Heckman et al. (\cite{HEC98}) and complemented by data from 
Leitherer et al. (\cite{LEI02}), Calzetti et al. (\cite{CAL94}), and Johnson 
et al. (\cite{JOH99}).}
\label{tab_IUE_basic}
\centering
\begin{tabular}{l l c c c}
\hline\hline
\noalign{\smallskip}
Name & Class & $z$ & $E_{B-V}$ & $O/H$ \\
& & & [mag] & \\
\noalign{\smallskip}
\hline
\noalign{\smallskip}
ESO\,338-4       & BCG             & 0.0096 & 0.09 & 8.1 \\
I\,Zw\,18        & BCDG            & 0.0025 & 0.03 & 7.2 \\
IRAS\,08339+6517 & H\,II           & 0.0191 & 0.09 & 8.7 \\
Mrk\,33          & BCDG            & 0.0049 & 0.01 & 8.4 \\
NGC\,1097        & Hs\,+\,Lin      & 0.0043 & 0.03 & 9.3 \\
NGC\,1313        & H\,II           & 0.0016 & 0.11 & 8.4 \\
NGC\,1510        & BCDG            & 0.0030 & 0.01 & 7.9 \\
NGC\,1705        & BCDG            & 0.0021 & 0.01 & 8.0 \\
NGC\,1741        & H\,II           & 0.0137 & 0.05 & 8.1 \\
NGC\,2782        & SB\,nuc.        & 0.0085 & 0.02 & 8.8 \\
NGC\,3049        & SB\,nuc.        & 0.0050 & 0.04 & 9.1 \\
NGC\,3353        & BCDG            & 0.0031 & 0.01 & 8.4 \\
NGC\,3738        & H\,II           & 0.0008 & 0.01 & 8.4 \\
NGC\,4214        & SB\,nuc.        & 0.0010 & 0.02 & 8.2 \\
NGC\,4385        & SB\,nuc.        & 0.0071 & 0.02 & 8.7 \\
NGC\,4449        & H\,II           & 0.0007 & 0.02 & 8.4 \\
NGC\,4670        & BCDG            & 0.0036 & 0.01 & 8.2 \\
NGC\,4861        & BCDG            & 0.0028 & 0.01 & 8.0 \\
NGC\,6052        & SB\,nuc.        & 0.0157 & 0.08 & 8.6 \\
NGC\,7496        & Sy\,2\,+\,H\,II & 0.0055 & 0.01 & 9.0 \\
NGC\,7673        & H\,II           & 0.0114 & 0.04 & 8.5 \\
NGC\,7714        & SB\,nuc.        & 0.0093 & 0.05 & 8.7 \\
NGC\,7793        & H\,II           & 0.0008 & 0.02 & 8.6 \\
UGC\,9560        & BCDG            & 0.0041 & 0.01 & 8.2 \\
\noalign{\smallskip}
\hline
\end{tabular}
\end{table}

In order to test our new parametric description of the rest-frame UV SED of 
star-forming galaxies, we select a sample of 24 well-known local starburst 
galaxies ($z \lesssim 0.02$; see Table~\ref{tab_IUE_basic}) with available UV 
spectra (see Kinney et al. \cite{KIN93}; Calzetti et al. \cite{CAL94}; 
Heckman et al. \cite{HEC98}). For this sample of nearby starbursts the 
luminosities at 1500\,\AA{} are typically one order of magnitude fainter than 
those of the FDF galaxies (cf. Leitherer et al. \cite{LEI02} and Noll et al. 
\cite{NOL04}). 

The UV spectra of this reference spectroscopic sample are taken from the INES 
archive of IUE newly-extracted spectra. The resolution element of the 
low-resolution IUE spectra is about 6\,\AA{}, i.e. similar to that achieved 
for the rest-frame UV spectra of the FDF sample of $z \sim 2$ UV-luminous 
galaxies (see Sect.~\ref{FDF}). A fine adjustment of the resolution of the 
IUE spectra was not needed, since this study focuses on the analysis of the 
UV continuum. Further selection criteria are as follows. First, all galaxies 
with a dominant young stellar population and available large 
$10'' \times 20''$ aperture, low-resolution UV spectra at short {\em and} 
long wavelengths were selected. In fact, only the combined spectrum obtained 
from the spectra taken with the short-wavelength prime camera (SWP) and the 
long-wavelength prime and redundant cameras (LWP and/or LWR) cover the 
wavelength range between 1150 and 3350\,\AA{}, bracketing the region of the 
2175\,\AA{} feature. Second, only those galaxies with spectra of suitable 
quality all over the UV spectral range were selected further. For obvious 
reasons, the data quality at the overlapping region of the two IUE spectral 
channels at around 2000\,\AA{} turned out to be a crucial parameter. Hence, 
combined spectra exhibiting a break in this overlapping region, usually with 
rather poor S/N, were rejected. As a result, the combined spectra of the 24 
selected IUE starbursts exhibit an average S/N of about 10 between 2000 and 
2350\,\AA{}, the final spectrum of each object being composed of all single 
spectra of sufficient S/N. In particular, the flux levels in the short and 
long wavelength spectra were adjusted in the wavelength range between 
1850\,\AA{} and 1980\,\AA{}. Finally, the combined spectra were corrected for 
Galactic extinction and mapped into the rest frame.

The correction for Galactic extinction makes use of the maps of Galactic dust
emission at IR wavelengths (Schlegel et al. \cite{SCHL98}) for the derivation
of realistic values for $E_{B-V}$, and of the \cite{CAR89} Milky Way 
extinction law for $R_{V} = 3.1$. This step is crucial, since the MW 
extinction law for the average ISM exhibits a relatively strong UV bump, so 
that small errors in $E_{B-V}$ can lead to important uncertainties in the 
evaluation of the UV bump strength from the spectra of the 24 selected IUE 
starbursts. However, for the sample galaxies, $E_{B-V}$ ranges between $0.01$ 
and $0.11$, with an average value of $0.03$ (see Table~\ref{tab_IUE_basic}), 
so that the effect of the correction for Galactic extinction on most of the 
spectra is possibly minimised.

\subsection{Basic properties of the FDF and IUE samples}\label{basics}

\begin{figure}
\centering 
\includegraphics[width=8.8cm,clip=true]{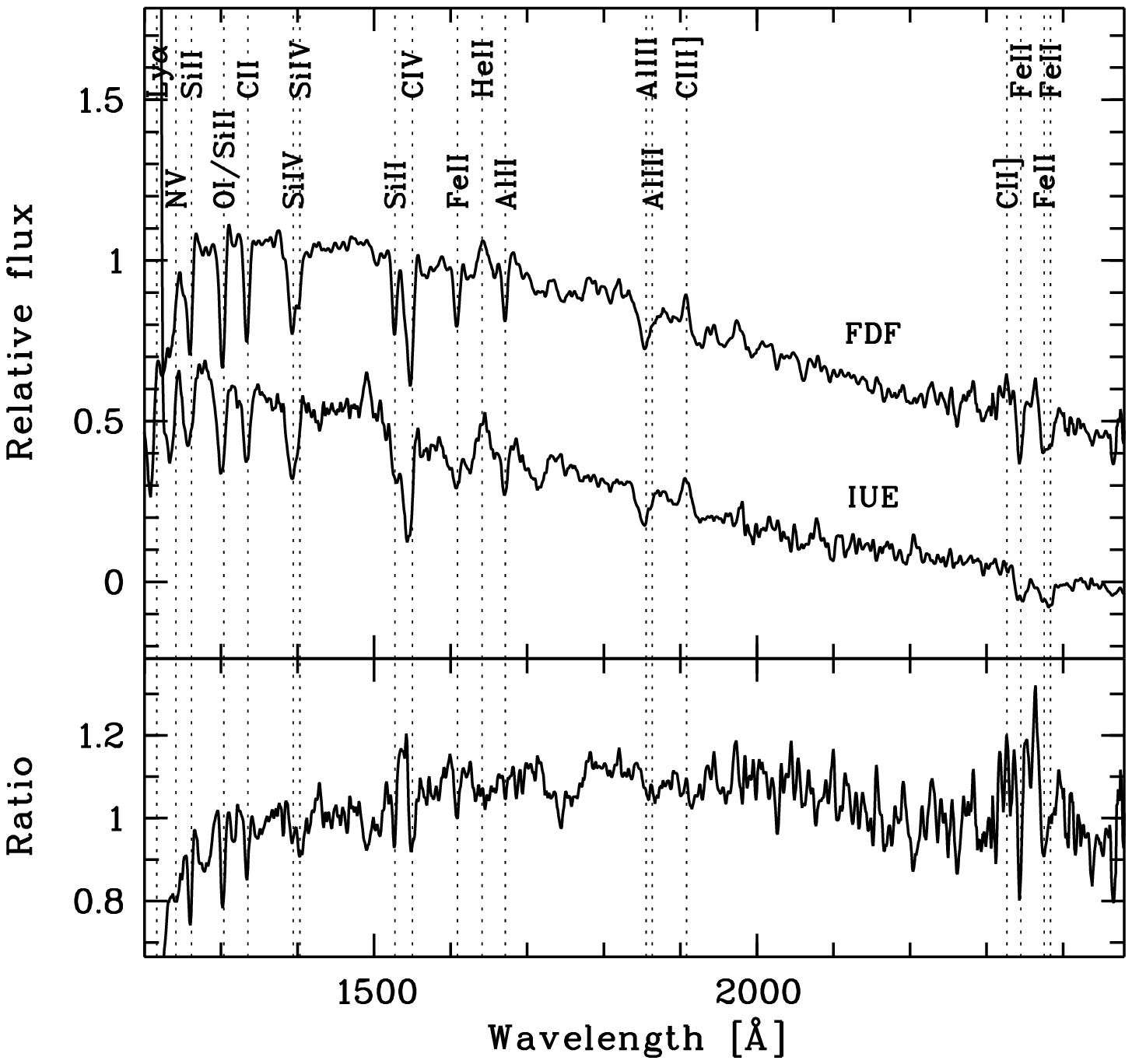}
\caption[]{Composite of 34 FDF galaxy spectra with S/N~$\ge 4$ in the range 
$2 < z < 2.5$ (upper spectrum in the upper panel) in comparison with a 
composite of 24 IUE spectra of local galaxies (lower spectrum in the upper 
panel; offset by $-0.5$ for clarity). The lower panel gives the ratio of the 
FDF and the IUE composite spectrum. The positions of prominent UV lines are 
indicated. The Ly$\alpha$ line of the IUE composite is strongly affected by 
emission from the geocorona.}
\label{fig_composites}
\end{figure}

\begin{table}
\caption[]{Properties of the composite spectra of the FDF $2 < z < 2.5$ and
the IUE local galaxy sample in comparison. The table lists the number of 
spectra averaged, the mean redshift, the mean equivalent width of six
low-ionisation interstellar absorption lines (see Noll et al. \cite{NOL04}),
and the EWs of C\,IV\,$\lambda\lambda\,1548,1550$, He\,II\,$\lambda\,1640$,
and C\,III]\,$\lambda\lambda\,1907,\,1909$. Emission lines have negative EWs. 
All errors given are mean errors.} 
\label{tab_ew}
\centering
\begin{tabular}{c c c}
\hline\hline
\noalign{\smallskip}
Sample & FDF & IUE \\
\noalign{\smallskip}
\hline
\noalign{\smallskip}
$N$ & $34$ & $24$ \\
$z$ & $2.32 \pm 0.02$ & $0.006 \pm 0.001$ \\
$W_{\rm LIS}$ [\AA] & $1.97 \pm 0.09$ & $1.89 \pm 0.08$ \\
$W_{\rm C\,IV}$ [\AA] & $3.80 \pm 0.16$ & $3.46 \pm 0.37$ \\
$W_{\rm He\,II}$ [\AA] & $-1.42 \pm 0.16$ & $-2.06 \pm 0.45$ \\
$W_{\rm C\,III]}$ [\AA] & $-1.09 \pm 0.18$ & $-1.91 \pm 0.23$ \\
\noalign{\smallskip}
\hline
\end{tabular}
\end{table}

Fig.~\ref{fig_composites} reproduces the composite spectrum of the 34 
UV-luminous FDF galaxies at $2 < z < 2.5$ and the composite spectrum of the 
24 local IUE starbursts. These composite spectra were calculated according to
the recipe given in Noll et al. (\cite{NOL04}). This means that the 
rest-frame spectra were first normalised to a common mode in the range 1250 
to 1500\,\AA{} and then averaged using equal weights. Wavelength regions 
affected by residuals of strong night sky lines or other defects were 
excluded from the averaging.

The lower panel of Fig.~\ref{fig_composites} shows the ratio of the FDF and 
IUE composites. In general, both rest-frame UV spectra are very similar, and 
show clear signatures of the dominance of young stellar populations. For a 
large wavelength range the continuum flux increases towards shorter 
wavelengths. However, for $\lambda < 1800$\,\AA{} the FDF composite is 
distinctly flatter than the IUE composite. Conversely, for 
$\lambda > 1800$\,\AA{} the local starburst composite shows a flatter slope 
than the other. We will discuss these behaviours in the following sections.

In addition to the overall UV continua, the average equivalent width of six
prominent low-ionisation lines ($W_{\rm LIS}$, see Shapley et al. 
\cite{SHA03}; Noll et al. \cite{NOL04}) are similar between the two samples 
(Fig.~\ref{fig_composites} and Table~\ref{tab_ew}). In fact, 
$W_{\rm LIS} = 1.97 \pm 0.09$\,\AA{} for the FDF composite and 
$1.89 \pm 0.08$\,\AA{} for the local one. The quoted errors include 
statistical uncertainties and the variance of the line strengths within each 
sample (see Noll et al. \cite{NOL04}). These low-ionisation lines originate 
in the intervening neutral interstellar medium between the OB stars 
(responsible of the rest-frame UV continuum) and the observer. Due to their 
high saturation, these lines are barely sensitive to chemical abundances. 
Hence, they primarily trace the spatial and velocity distribution of the 
neutral interstellar gas illuminated by the rest-frame UV continuum. 
Therefore, $W_{\rm LIS}$ is a measure of the product of the neutral clouds' 
covering fraction and the velocity spread. According to Shapley et al.
(\cite{SHA03}) differences in $W_{\rm LIS}$ are mainly due to variations in
the covering fraction.

Comparable equivalent widths are also obtained for the important 
high-ionisation C\,IV doublet: $3.80 \pm 0.14$\,\AA{} for the FDF sample,
and $3.46 \pm 0.37$\,\AA{} for the IUE sample. The C\,IV equivalent width
mainly measures the strength of winds from hot, massive and, thus, young 
stars and is a good metallicity indicator (Leitherer et al. \cite{LEI01}). 
Using C\,IV Mehlert et al. (\cite{MEH02})\footnote{The EWs of C\,IV listed in
Mehlert et al. (\cite{MEH02}) are on average about 20\% larger than in this 
study. This is due to a wider wavelength range to derive the continuum level 
(190\,\AA{} instead of $\sim 80$\,\AA{}) and an automatic instead of 
interactive measurement technique.} estimated an average metallicity of 
$Z \approx 0.42$\,Z$_{\odot}$ for an analogous sample of FDF galaxies at 
$2 < z < 2.5$. Measuring the purely photometric indices ``1370'' and ``1425'' 
(Leitherer et al. \cite{LEI01}) in medium-resolution spectra of two 
$z \sim 2$ FDF galaxies Mehlert et al. (\cite{MEH05}) find indications for 
even higher metallicities close to the solar value. A similar metallicity as 
the C\,IV based of the FDF galaxies (assuming the solar oxygen abundance 
$(O/H)_{\odot} = 8.93$) we find for our IUE sample of nearby starbursts 
having $\langle O/H \rangle \approx 8.6$ (see Table~\ref{tab_IUE_basic} and, 
e.g., Heckman et al. \cite{HEC98}). 

In addition to the P~Cygni profiles of high-ionisation lines such as C\,IV,
the detection of the He\,II and C\,III] emission features indicates the 
presence of a significant population of young, very massive stars in both 
samples (Fig.~\ref{fig_composites} and Table~\ref{tab_ew}). In particular, 
the He\,II line emission is essentially produced by Wolf-Rayet stars 
(Schaerer \cite{SCHA03}), whereas the C\,III] line emission has a nebular 
origin, and traces the electron temperature of H\,II regions (see Heckman et 
al. \cite{HEC98}).

\section{Analysis}\label{analysis}

In this study we compare diagnostics measured on the rest-frame UV spectra
of the 34 FDF sample galaxies (see Sect.~\ref{FDF}) with those determined
on the rest-frame UV SEDs of models of starburst galaxies. This procedure
is analogous to that adopted by Gordon et al. (\cite{GOR97}) for the analysis
of the dust properties for the Calzetti et al. (\cite{CAL94}) sample of 
nearby starbursts. In addition, we use the IUE sample of 24 local starbursts
(see Sect.~\ref{IUE}) as a further comparison. Here we illustrate both the 
models and the diagnostics adopted in this study. Furthermore, we demonstrate 
that the new diagnostics tailored to the FORS data lead to the same 
conclusions as those of Gordon et al. (\cite{GOR97}) when applied to the IUE 
sample.

\subsection{Modelling the UV continuum of starburst galaxies}\label{models}

We produce synthetic SEDs for starburst galaxies by combining stellar 
population synthesis models with models of radiative transfer of the stellar 
and scattered radiation through the dusty ISM.

\subsubsection{Stellar population synthesis models}\label{SPS}

Intrinsic (i.e. non attenuated by dust) SEDs for starburst galaxies are 
derived from the stellar population synthesis code for the stellar 
continuum emission of Maraston (2005, \cite{MAR05})\footnote{The Maraston 
(2005) models are available at 
\texttt{http://www-astro.physics.ox.ac.uk/$\sim$maraston/}.}. These models 
include the thermally pulsating asymptotic giant branch phase for 
intermediate-mass ($2 - 5$\,M$_{\odot}$) stars, at variance with all others 
(e.g. Fioc \& Rocca-Volmerange 1997, \cite{FIO97}; Leitherer et al. 
\cite{LEI99}; Bruzual \& Charlot 2003, \cite{BRU03}). However, this relevant 
difference does not affect any of the results reported here, since the SEDs 
produced for analogous models of \cite{MAR05}, \cite{BRU03}, and \cite{FIO97}
do not differ significantly in shape at rest-frame UV wavelengths (see, e.g., 
Fig.~14 in \cite{MAR05}). This is also shown in Table~\ref{tab_dgamma_sps} of 
Sect.~\ref{sensitivity}. Finally, we note that the spectral resolution of the
\cite{MAR05} models ($\Delta\lambda = 10$\,\AA{}) matches reasonably well 
those of the rest-frame UV continua of the FDF sample galaxies, as given by 
the FORS spectra ($\Delta\lambda = 5 - 8$\,\AA{}), and of the nearby 
starbursts observed with IUE ($\Delta\lambda = 6$\,\AA{}).

In Sect.~\ref{basics}, we have shown that there is evidence for the 
$z \sim 2$ UV-luminous FDF galaxies being both actively forming stars and 
metal enriched ($\langle Z \rangle \gtrsim 0.4$\,Z$_{\odot}$). In addition,
starburst galaxies in the nearby and distant universe seem to undergo multiple 
star-formation episodes of a few 10\,Myr to several 100\,Myr (Calzetti 
\cite{CAL97}; Shapley et al. \cite{SHA01}; van Dokkum et al. \cite{DOK04}). 
Therefore, we adopt composite stellar populations models where the SFR is 
constant with time, the metallicity is fixed and ranges between 2\% and twice 
solar, and the time elapsed since the start of the star-formation activity 
(i.e. the age of the model) ranges between 5 and 500\,Myr. As a reference, we 
compute additional simple stellar population models spanning the same 
metallicity range as for the previous models but with ages between 5 and 
100\,Myr. All these models assume a Salpeter (\cite{SAL55}) initial mass 
function with an upper mass cut-off of 100\,M$_{\odot}$. 

Given the measured metallicities (Mehlert et al. \cite{MEH02}, \cite{MEH05}) 
and the expected ages of the $z \sim 2$ UV-luminous galaxies, in the following 
analysis we will make use mostly of \cite{MAR05} models with a continuous 
star-formation activity (i.e. with a SFR $e$-folding time $\tau = 10$\,Gyr) 
along 50, 200, or 500\,Myr, and metallicities equal to $0.5$ or 
$1.0$\,Z$_{\odot}$. These models are also suitable to reproduce the 
properties of the IUE sample of nearby starbursts, though a few objects in 
the latter sample have $Z < 0.1$\,Z$_{\odot}$ (e.g. I\,Zw\,18) and a few 
others have significantly supersolar metallicity (see 
Table~\ref{tab_IUE_basic}).

\subsubsection{Radiative transfer models}\label{radtransfer}

\begin{figure}
\centering 
\includegraphics[width=8.8cm,clip=true]{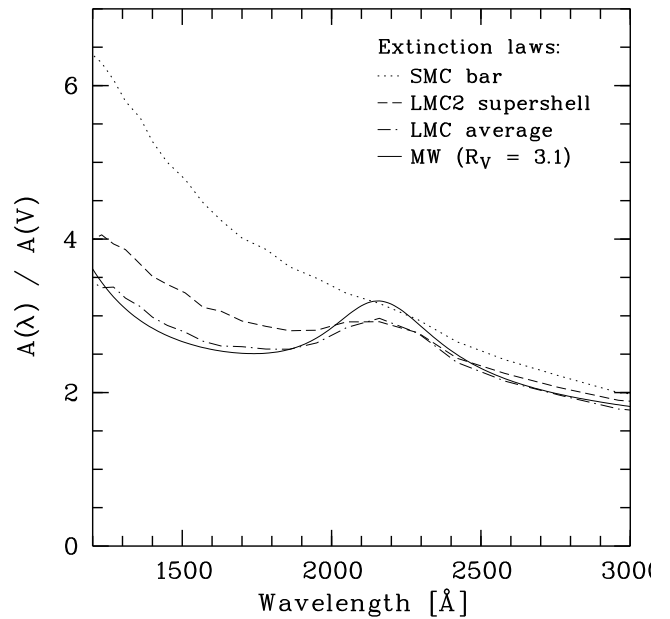}
\caption[]{Ratio of the attenuation factor $A$ at the wavelength $\lambda$ to 
the average attenuation in the $V$ band as function of $\lambda$ for 
different extinction laws. The \cite{CAR89} Milky Way extinction law is shown 
for $R_V = A_V / E_{B-V} = 3.1$ as solid line. The LMC average, LMC\,2 
supershell, and SMC bar extinction laws (Gordon et al. \cite{GOR03}) are 
marked by dash-dotted, dashed, and dotted lines, respectively.}
\label{fig_extinct}
\end{figure}

\begin{figure}
\centering 
\includegraphics[width=8.8cm,clip=true]{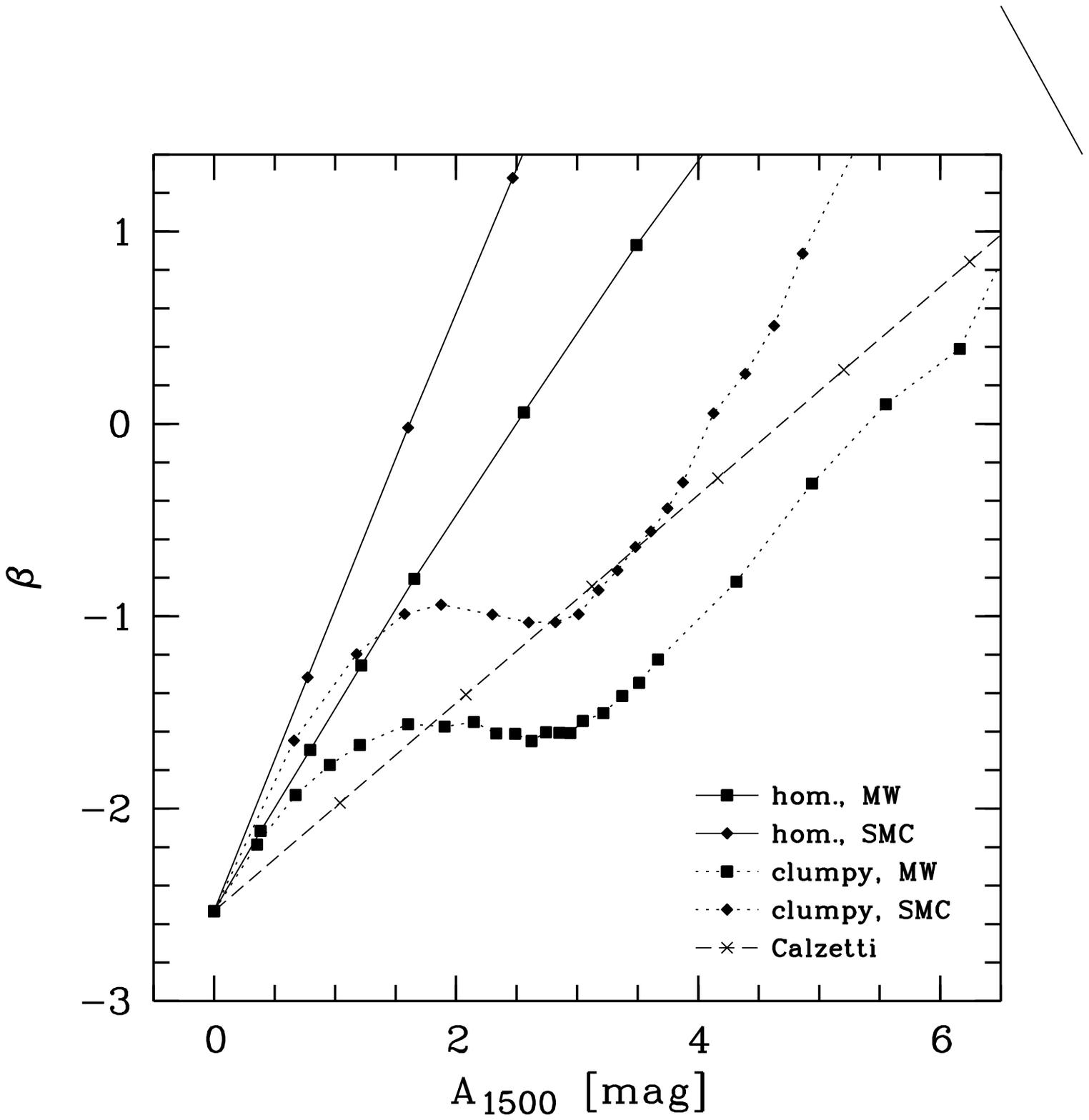}
\caption[]{The UV continuum slope $\beta$ as a function of the attenuation 
factor at 1500\,\AA{} $A_{1500}$ for different reddened model spectra. In all
cases a \cite{MAR05} model with Salpeter IMF, continuous star-formation, an 
age of 200\,Myr, and solar metallicity was combined with SHELL models of 
\cite{WIT00} for different dust distribution, extinction laws, and optical 
depth. Different dust column densities are indicated by filled squares (MW 
dust) and lozenges (SMC dust), connected with solid (homogeneous model) and 
dashed lines (clumpy model), respectively. The symbols have been plotted for 
the average radial visual optical depth 
$\tau_V = 0, 0.25, ..., 1, 1.5, ..., 6, 7, ..., 10$. The dash-dotted line 
refers to the Calzetti et al. (\cite{CAL00}) attenuation law. The crosses are 
plotted in intervals of $\Delta E_{B-V} = 0.1$.}
\label{fig_beta_A1500}
\end{figure}

We make use of the radiative transfer calculations for the SHELL dust/stars 
configuration of \cite{WIT00} in order to describe the properties of dust 
attenuation for the galaxies under investigation. The reason is that the 
rather regular properties of dust attenuation in starburst galaxies of the 
local universe (Calzetti et al. \cite{CAL94}, \cite{CAL00}) may be reproduced 
by a model where a two-phase, clumpy, dusty medium is distributed in a shell 
surrounding the stars (Gordon et al. \cite{GOR97}). This dust/stars 
configuration is analogous to that adopted by Fischera et al. (\cite{FIS03}) 
to give an independent physical explanation of the so-called ``Calzetti law'' 
(Calzetti et al. \cite{CAL94}, \cite{CAL00}) based on a turbulent ISM. 
According to Dopita et al. (\cite{DOP05}), the pressure in the diffuse phase 
of the ISM (or, equivalently, its density), and the molecular-cloud 
dissipation timescale control the pan-spectral SED of starburst galaxies.

The \cite{WIT00} radiative transfer models (including multiple scattering) 
are based on Monte Carlo calculations for a scale-free spherically-symmetric 
geometry (see also Gordon et al. \cite{GOR01}). In particular, the SHELL 
dust/stars configuration assumes the stars to be distributed inside an inner 
sphere with radius equal to $0.3$ times the system radius, and the dust to be 
distributed inside a shell extending radially from $0.3$ to 1 times the 
system radius. Dust can be distributed locally either in a homogeneous (i.e. 
diffuse) ISM, or in a two-phase, clumpy ISM. In the latter case, high-density 
clumps are embedded in a low-density interclump medium, the 
low-to-high-density ratio being equal to $0.01$, with a filling factor of 
$0.15$. Models are parameterised by the radial extinction optical depth from 
the centre to the edge of the dust environment at $V$ band ($\tau_V$), 
assuming a constant density homogeneous distribution, $\tau_V$ ranging from 
$0.25$ to 50. Finally, the mixture of dust grains can be characterised by an 
extinction curve that is typical either of the SMC or of the MW (see 
Fig.~\ref{fig_extinct} and Sect.~\ref{introduction}).

Hereafter, we illustrate a few aspects of these SHELL radiative transfer 
models for the non-specialist reader. Fig.~\ref{fig_beta_A1500} shows the 
behaviour of the UV spectral slope $\beta$\footnote{The non-ionising 
UV-flux density distribution of an observed star-forming galaxy is well 
reproduced by a power law $f_{\lambda} \sim \lambda^{\beta}$, where $\beta$ 
is referred to as the UV slope (see, e.g., Calzetti et al. \cite{CAL94}).},
as defined in Sect.~\ref{technique}, as a function of the attenuation (in mag) 
at 1500\,\AA{} $A_{1500}$ for different starburst models with age equal to 
200\,Myr and solar metallicity (cf. Sect.~\ref{SPS}). The relation between 
$\beta$ and $A_{1500}$ is linear in case of a homogeneous dusty ISM, whatever 
the dust type. Conversely, it is non-linear for the two-phase, clumpy dust 
distribution. In the former case, $\beta$ and $A_{1500}$ increase at a 
constant rate because the angle-averaged optical depth at UV wavelengths 
increases monotonically with the dust column density (i.e. $\tau_V$). In the 
latter case, first $\beta$ and $A_{1500}$ show a positive correlation 
because the optical depth at UV wavelengths of a clump increases with 
$\tau_V$. However, when the clumps become optically thick at UV wavelengths, 
the attenuation at these wavelengths is given by the simple blocking action 
of the clumps, while the UV radiation reaches the outside observer through 
the optically thin, diffuse component of the ISM. Hence, $\beta$ stalls while 
$A_{1500}$ keeps on increasing. When the angle-averaged optical depth at UV 
wavelengths of the diffuse component of the ISM tends to unity, $\beta$ and 
$A_{1500}$ increase together again. We note that, in case of a more realistic 
size and density distribution of the dusty clumps, one can expect a behaviour 
that is intermediate between those produced by the homogeneous and two-phase, 
clumpy dust distributions. Furthermore, Fig.~\ref{fig_beta_A1500} confirms 
that SMC-type dust produces a larger attenuation at UV wavelengths than 
MW-type dust, for fixed $\tau_V$, whatever the structure of the dusty ISM.

Finally, Fig.~\ref{fig_beta_A1500} shows that a linear relation between 
$\beta$ and $A_{1500}$ is expected when the empirical ``Calzetti law'' is 
assumed to hold, whatever the dust column density of the ISM. We remind that 
\cite{WIT00} could achieve the best agreement with the ``Calzetti law'' for a 
two-phase, clumpy SHELL model with SMC dust, and $\tau_V = 1.5$.

\subsection{A new parametric description of the UV continuum}\label{method}

\subsubsection{The parameter set}\label{technique}

\begin{table}
\caption[]{Wavelength windows for the derivation of the UV continuum slope 
parameters $\beta$ and $\gamma_1$ to $\gamma_4$ using power-law fits}
\label{tab_windows}
\centering
\begin{tabular}{c l}
\hline\hline
\noalign{\smallskip}
Par. & Windows \\
\noalign{\smallskip}
\hline
\noalign{\smallskip}
$\beta$    & $1268-1284$, $1309-1316$, $1342-1371$, $1435-1496$, \\
           & $1562-1583$, $1677-1740$ \\
$\gamma_1$ & $1268-1284$, $1309-1316$, $1342-1371$, $1435-1496$ \\
$\gamma_2$ & $1562-1583$, $1677-1740$, $1760-1833$, $1866-1890$ \\
$\gamma_3$ & $1920-2050$, $2070-2175$ \\
$\gamma_4$ & $2175-2310$, $2400-2480$ \\
\noalign{\smallskip}
\hline
\end{tabular}
\end{table}

\begin{figure}
\centering 
\includegraphics[width=6cm,angle=-90]{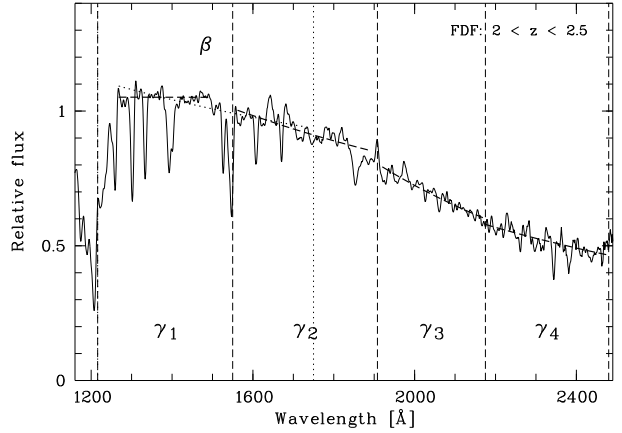}
\caption[]{Demonstration of the derivation of the UV continuum slopes $\beta$ 
and $\gamma_1$ to $\gamma_4$ for the composite spectrum of 34 luminous FDF
galaxies at $2 < z < 2.5$ (cf. Fig.~\ref{fig_composites}). The power law fits 
are shown as thick dotted ($\beta$) and dashed ($\gamma_1$ to $\gamma_4$)
curves. The limits for the fit ranges are indicated by vertical dotted 
($\beta$) and dashed ($\gamma_1$ to $\gamma_4$) lines.}
\label{fig_demogamma}
\end{figure}

The UV SEDs of composite stellar population models can be parameterised with 
a power law ($f(\lambda) \propto \lambda^{\beta}$), since the bulk radiation 
comes from the emission from hot stars in the Rayleigh-Jeans regime. This is 
so also true for observed UV SEDs of nearby starbursts (e.g. Calzetti et al. 
\cite{CAL94}), that are affected by dust attenuation (see footnote~3). The 
latter authors measured the UV slope $\beta$ across the wavelength range 
$1200 - 2600$\,\AA{}, with the exclusion of the wavelength region between 
1950 and 2400\,\AA{}. By doing so, their $\beta$ was not affected by the 
possible presence of dust characterised by an extinction law with the UV bump. 
Other analyses of the UV continuum properties of star-forming galaxies in the 
local and distant universe defined $\beta$ across a narrower range, i.e. from 
1200 to 1800\,\AA{} (Heckman et al. \cite{HEC98}; Leitherer et al. 
\cite{LEI02}; Noll et al. \cite{NOL04}), mostly owing to the missing spectral 
coverage at rest-frame $\lambda > 2400$\,\AA{}. Following the previous authors
we also measure $\beta$ across the wavelength range $1200 - 1800$\,\AA{}. The
quality of the FORS spectra available for the 34 FDF galaxies at $2 < z < 2.5$
is not good enough to provide a reliable fit at $\lambda > 2400$\,\AA{}.

In addition to the ``standard'' parameter $\beta$, we define six new 
parameters, since the main interest of this study is to investigate if and 
when the UV bump is present among the FDF sample galaxies. The definition of 
four parameters (see Table~\ref{tab_windows}) is based on power-law fits to 
sub-regions of the UV continuum as well as that of $\beta$. In particular, we 
divide the wavelength range between 1200 and 2500\,\AA{} in four sub-regions, 
delimited by the C\,IV doublet at 1550\,\AA{}, the C\,III] emission feature 
at 1908\,\AA{}, and the centre of the UV bump at 2175\,\AA{}. Furthermore, we 
select windows within a given sub-region (see Table~\ref{tab_windows}) in 
order to avoid that the presence of strong spectral lines produces a bias of 
the fit to the continuum. For $\lambda < 1800$\,\AA{} we essentially adopt 
the windows defined by Calzetti et al. (\cite{CAL94}), with the exception of 
the $1435 - 1496$\,\AA{} window, which is taken from Leitherer et al. 
(\cite{LEI02}). We call these four parameters $\gamma_1$, $\gamma_2$, 
$\gamma_3$, and $\gamma_4$ after the progressive number of the four 
sub-regions, starting from that defined by the shortest wavelengths. 
Fig.~\ref{fig_demogamma} reproduces the five power-law fits corresponding to 
the previous five parameters (i.e. $\beta$ plus the four new slopes of the 
fits) for the composite spectrum of 34 UV-luminous FDF galaxies at 
$z \sim 2$. In addition, we introduce two parameters, i.e. 
$\gamma_{12} = \gamma_1 - \gamma_2$ and $\gamma_{34} = \gamma_3 - \gamma_4$. 
The meaning of these two additional parameters is straightforward: 
$\gamma_{12}$ measures the ''curvature'' of the spectral UV continuum across 
1500\,\AA{}; $\gamma_{34}$ measures the ``curvature'' of the spectral UV 
continuum across the UV bump (negative $\gamma_{34}$ identify the presence of 
dust with an extinction law exhibiting the 2175\,\AA{} feature). The 
parameter $\gamma_{34}$ does not represent the ``strength'' of the UV bump 
however. In fact, radiative transfer effects tend to dilute the prominence of 
the UV bump in the resulting attenuation function with respect to that in the 
extinction curve when the opacity increases, whatever the dust/stars 
configuration and the structure of the dusty ISM, with the exception of a 
homogeneous SHELL configuration (Cimatti et al. \cite{CIM97}; \cite{WIT00}; 
Pierini et al. \cite{PIE04}).

\subsubsection{Sensitivity of the parameters to properties of the stellar 
populations or of dust attenuation}\label{sensitivity}

\begin{figure}
\centering 
\includegraphics[width=8.8cm,clip=true]{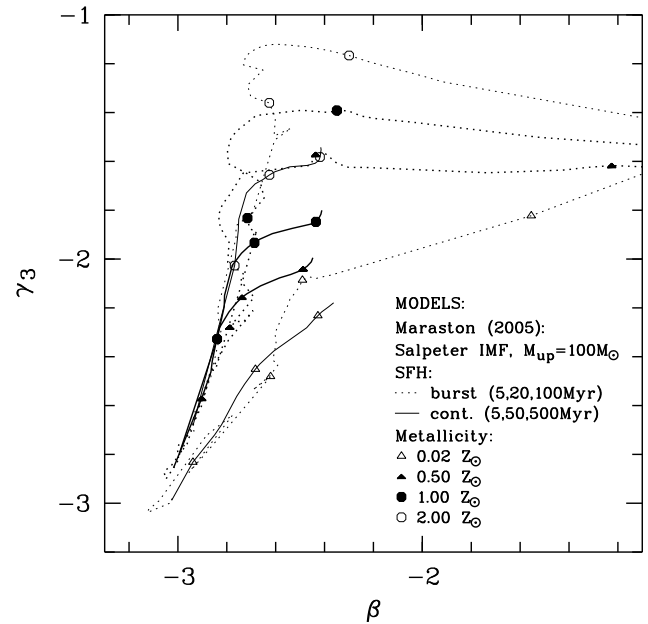}
\caption[]{The relation between the slope parameters $\gamma_3$ and $\beta$ 
for different \cite{MAR05} model spectra. The tracks plotted show the change 
of $\beta$ and $\gamma_3$ with increasing age ($t \le 1$\,Gyr) for stellar 
population models differing in the star-formation history (single burst or 
continuous star-formation) as well as the metallicity ($0.02$, $0.5$, $1.0$, 
and $2.0$\,Z$_{\odot}$). For single burst models (dotted lines) symbols (see 
legend) are plotted for 5, 20, and 100\,Myr. Ages of 5, 50, and 500\,Myr are 
marked in case of models with continuous star-formation (solid lines). All 
tracks start at $\beta \sim -3$ and $\gamma_3 \sim -3$.}
\label{fig_gambet_sps}
\end{figure}

\begin{table}
\caption[]{Continuum slope parameters for continuous star-formation models of 
\cite{MAR05}, differing in metallicity ($0.5$ and 1\,Z$_{\odot}$) and age (50 
and 500\,Myr), and for analogous models of \cite{BRU03} and \cite{FIO97}}
\label{tab_dgamma_sps}
\centering
\begin{tabular}{c c c c c c}
\hline\hline
\noalign{\smallskip}
Par. & $0.5$\,Z$_{\odot}$ & 1\,Z$_{\odot}$ & $0.5$\,Z$_{\odot}$ & 
1\,Z$_{\odot}$ & Ref. \\
\noalign{\smallskip}
\hline
\noalign{\smallskip}
$\beta$    & $-2.74$ & $-2.69$ & $-2.49$ & $-2.44$ & M05 \\
$\gamma_1$ & $-2.80$ & $-2.87$ & $-2.41$ & $-2.47$ & M05 \\
$\gamma_2$ & $-2.23$ & $-1.98$ & $-2.00$ & $-1.74$ & M05 \\
$\gamma_3$ & $-2.16$ & $-1.93$ & $-2.04$ & $-1.85$ & M05 \\
$\gamma_4$ & $-3.04$ & $-3.03$ & $-2.92$ & $-2.96$ & M05 \\
\noalign{\smallskip}
\hline
\noalign{\smallskip}
$\beta$    & $-2.73$ & $-2.67$ & $-2.45$ & $-2.39$ & BC03 \\
$\gamma_1$ & $-2.83$ & $-2.96$ & $-2.40$ & $-2.50$ & BC03 \\
$\gamma_2$ & $-2.23$ & $-1.86$ & $-1.97$ & $-1.60$ & BC03 \\
$\gamma_3$ & $-2.07$ & $-1.75$ & $-1.95$ & $-1.68$ & BC03 \\
$\gamma_4$ & $-3.00$ & $-3.01$ & $-2.88$ & $-2.95$ & BC03 \\
\noalign{\smallskip}
\hline
\noalign{\smallskip}
$\beta$    & $-2.61$ & $-2.58$ & $-2.36$ & $-2.33$ & FR97 \\
$\gamma_1$ & $-2.58$ & $-2.70$ & $-2.19$ & $-2.31$ & FR97 \\
$\gamma_2$ & $-2.24$ & $-1.94$ & $-2.01$ & $-1.71$ & FR97 \\
$\gamma_3$ & $-2.11$ & $-1.85$ & $-2.00$ & $-1.79$ & FR97 \\
$\gamma_4$ & $-2.85$ & $-2.88$ & $-2.76$ & $-2.84$ & FR97 \\
\noalign{\smallskip}
\hline
\end{tabular}
\end{table}

\begin{figure}
\centering 
\includegraphics[width=8.8cm,clip=true]{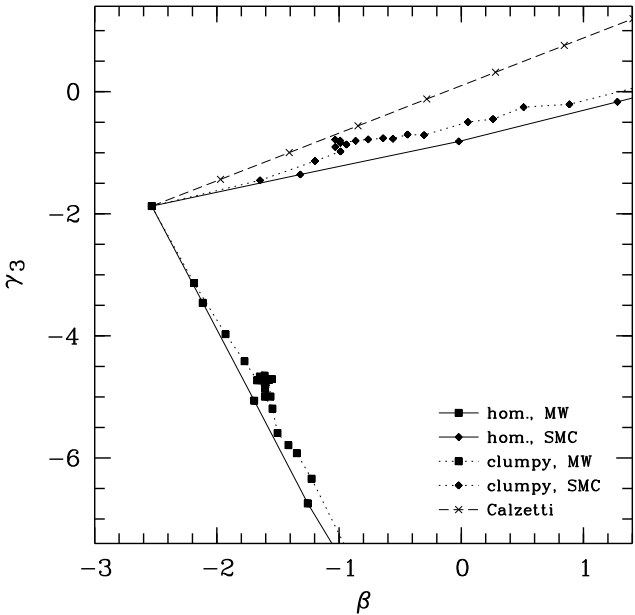}
\caption[]{The relation between the slope parameters $\gamma_3$ and $\beta$ 
for different reddened model spectra. The line types and symbols are as in 
Fig.~\ref{fig_beta_A1500}.}
\label{fig_gambet_dust}
\end{figure}

The five parameters $\beta$, and $\gamma_1$ to $\gamma_4$ describe the 
observed UV SEDs of a galaxy and, thus, depend in a complex way on the 
properties of the stellar populations and of dust attenuation (see 
Sect.~\ref{models}). Nevertheless, we can disentangle the principal 
dependencies of each parameter by investigating the behaviour of the five 
slopes, as obtained for the UV SEDs of a suite of starburst models, as a 
function of the model parameters.

In Fig.~\ref{fig_gambet_sps} we illustrate the effects on the UV continuum
parameters $\beta$ and $\gamma_3$ of different metallicities and ages, for 
single-burst and continuous star-formation models of \cite{MAR05}. A priori, 
one would expect that $\gamma_3$ is particularly sensitive to very young ages 
and high metallicities, since it characterises the wavelength range from 1900 
to 2175\,\AA{}. However, $\gamma_3$ hardly changes for reasonable models of 
$z \sim 2$ UV-luminous galaxies, with continuous star-formation, metallicity 
equal to $0.5$ or $1.0$\,Z$_{\odot}$, and ages higher than a few 10\,Myr (see 
Sect.~\ref{SPS}). The same conclusion holds for $\beta$: conspicuous changes 
in $\beta$ would be possible only for single burst models with ages greater 
than 20\,Myr. The latter models are not realistic, since they do not provide 
large UV luminosities long after the episode of star-formation activity. 
Furthermore, values of $\beta > -2$ are produced by stellar population 
synthesis models with a SFR $e$-folding time as short as 1\,Gyr only about 
10\,Gyr after the onset of the star-formation activity, i.e. at a time much 
longer than the age of the universe at $z \sim 2$ (i.e. $\sim 3$\,Gyr). SFR 
$e$-folding times less than 1\,Gyr can not justify the vigorous star-formation 
activity of the UV-luminous FDF galaxies at $z \sim 2$ that is witnessed by 
the measured emission line EWs (see Table~\ref{tab_ew}).

Table~\ref{tab_dgamma_sps} lists typical values (see Sect.~\ref{SPS}) for all 
five UV continuum slope parameters. In particular, it shows that $\gamma_4$ 
exhibits only a minor dependence on the properties of the stellar 
populations, while $\gamma_2$ is particularly sensitive to changes in 
metallicity {\em and} age. The latter is explained by the great number of 
spectral lines originated in the atmospheres of hot stars that fall within 
the wavelength range 1550 to 1900\,\AA{}, where $\gamma_2$ is defined, and 
the sensitivity of the same wavelength range to the O-to-B-star number ratio. 
Finally, Table~\ref{tab_dgamma_sps} shows that the slope parameters span a 
range $\Delta\gamma$ of up to $0.5$ in response to the variations in the 
properties of the stellar populations that are assumed to be reasonable for 
our sample galaxies. On the other hand, Table~\ref{tab_dgamma_sps} shows that 
the different $\gamma$ parameters characterising the rest-frame UV SED can 
change by up to $0.2$ with respect to those obtained from the \cite{MAR05} 
models, when they are determined from intrinsic rest-frame UV SEDs taken from 
\cite{BRU03} or \cite{FIO97}. Hence, the dependence of all the seven UV 
continuum slope parameters on the source of the stellar population synthesis 
code does not affect the study of the extinction law in UV luminous galaxies 
at $z \sim 2$.

Now, stellar population synthesis models do not include the absorption of UV 
light by interstellar gas, so that one may expect that the parameters listed 
in Table~\ref{tab_dgamma_sps} are affected by systematic uncertainties.
Therefore, we have computed semi-analytic spectra including line absorption
by the ISM by combining the normalised, composite rest-frame UV spectrum of 
our high-redshift sample galaxies (Fig.~\ref{fig_composites}) with the UV 
spectra of the \cite{MAR05} models listed in Table~\ref{tab_dgamma_sps}. As a 
result, we estimate that the deviations in the slope parameters listed there, 
as produced by interstellar line-absorption, are typically about $0.1$. This 
very small effect is also a consequence of the careful selection of 
wavelength windows excluding strong interstellar lines, where the five UV 
continuum slope parameters are defined (see Fig.~\ref{fig_composites}).

Finally, we discuss how the latter parameters depend on the properties of 
dust attenuation, for the models discussed in Sect.~\ref{radtransfer}. In 
particular, we illustrate the behaviour of dusty, star-forming galaxies with 
a solar metallicity, 200\,Myr-old composite stellar population in the 
$\beta - \gamma_3$ plane (see Fig.~\ref{fig_gambet_dust}) already introduced 
(see Fig.~\ref{fig_gambet_sps}). In this diagram, the loci of models with MW 
and SMC dust are clearly separated, whatever $\tau_V$. The slope parameters 
$\gamma_3$ and $\beta$ of the two models differ by more than $1.6$ and $0.5$, 
respectively, already for $\tau_V \gtrsim 0.25$. Furthermore, 
$\gamma_3 \sim -2$ is a delimiter of the regions populated by the models for
MW-like and SMC-like dust. Hence, the distribution in the $\beta - \gamma_3$ 
plane can reveal the presence of dust with an extinction law exhibiting the 
UV bump already for low values of the dust column density. On the other hand,
it is not particularly sensitive to the structure of the dusty ISM. As a 
reference, Fig.~\ref{fig_gambet_dust} reproduces additional models with the 
same stellar populations as the others but where dust attenuation follows the 
Calzetti et al. (\cite{CAL00}) attenuation law. As expected (Gordon et al. 
\cite{GOR97}), the latter models show a similar behaviour to those with a 
SHELL configuration of SMC-type dust distributed in a two-phase, clumpy 
medium.

In conclusion, the five UV continuum slopes defined in this study are proved 
to be much more sensitive to the characteristic extinction law (i.e. to the 
properties of the mixture of the dust grains) of the system, than to those of 
the stellar populations present in the system, for a fixed dust/stars 
distribution. Conversely, they cannot constrain the structure of the dusty 
ISM.

\subsubsection{A test for the method: the IUE sample of nearby starbursts}
\label{test}

\begin{figure}
\centering
\includegraphics[width=8.8cm,clip=true]{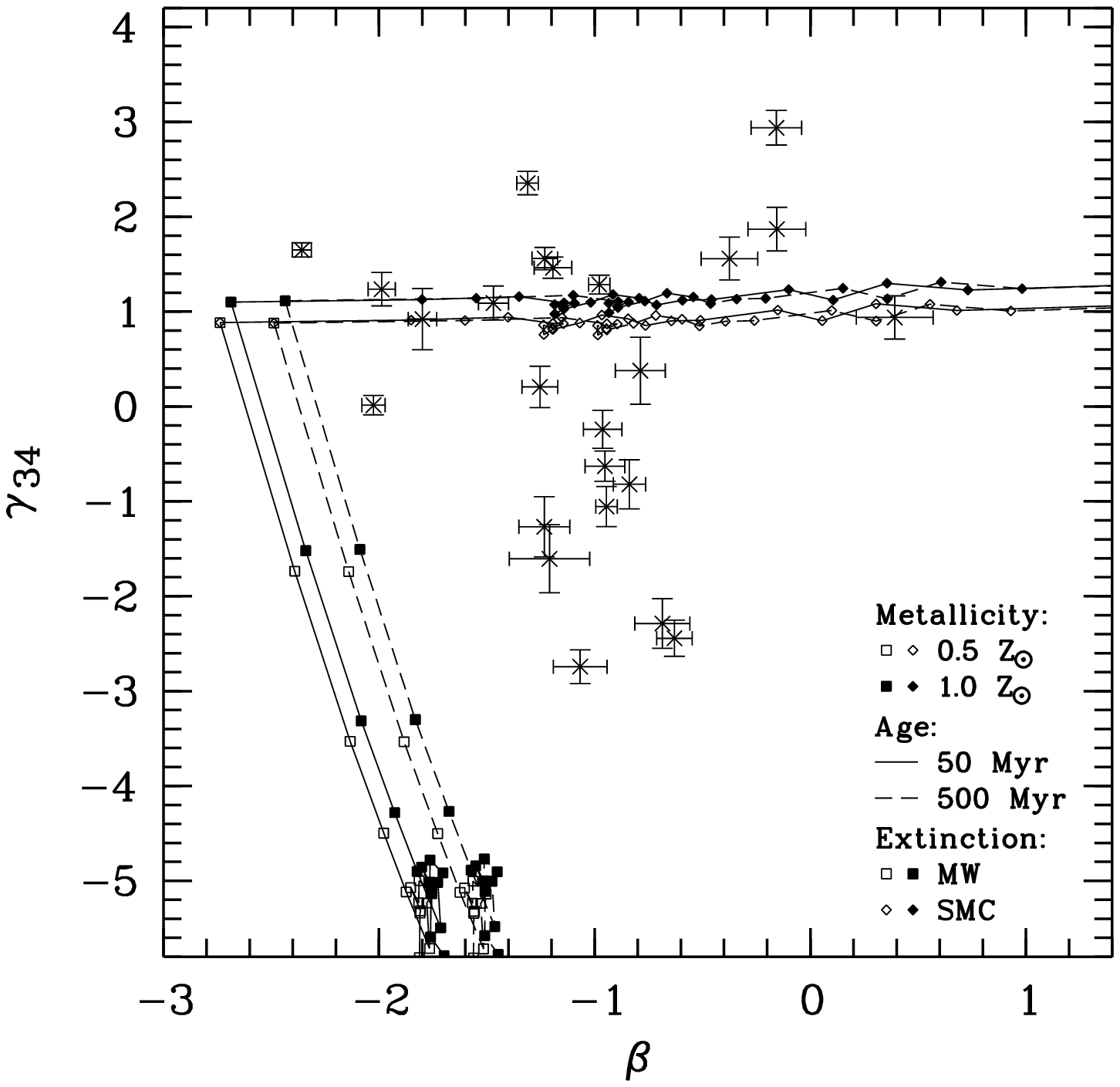}
\caption[]{The combined slope parameter $\gamma_{34}$ versus $\beta$ for the 
IUE sample of local starburst galaxies. The statistical uncertainties of the 
slope measurements are indicated by error bars. For $\gamma_{34}$ these are
probably smaller than the systematic uncertainties (see Section~\ref{IUE}).
The spectroscopic data are compared to combined \cite{MAR05} stellar 
population synthesis and \cite{WIT00} radiative transfer models. All 
\cite{MAR05} models shown (see legend) were calculated for a standard 
Salpeter IMF and continous star-formation. All \cite{WIT00} models given (see 
legend) were calculated for a clumpy SHELL dust/stars configuration. The free 
parameter of the curves is the average radial visual optical depth $\tau_V$ 
(see Fig.~\ref{fig_beta_A1500}).}
\label{fig_gambet_test} 
\end{figure}

\begin{table*}
\caption[]{UV continuum slope parameters of our IUE sample of 24 local 
starburst galaxies (see Table~\ref{tab_IUE_basic})}
\label{tab_IUE_gamma}
\centering
\begin{tabular}{l c c c c c c c}
\hline\hline
\noalign{\smallskip}
Name & $\beta$ & $\gamma_1$ & $\gamma_2$ & 
$\gamma_3$ & $\gamma_4$ & $\gamma_{12}$ & $\gamma_{34}$ \\
\noalign{\smallskip}
\hline
\noalign{\smallskip}
ESO\,338-4       & $-1.19 \pm 0.09$ & $+0.02 \pm 0.08$ & $-1.06 \pm 0.07$ & 
$-1.55 \pm 0.09$ & $-3.01 \pm 0.07$ & $+1.08 \pm 0.11$ & $+1.46 \pm 0.11$ \\
I\,Zw\,18        & $-1.80 \pm 0.07$ & $-1.20 \pm 0.06$ & $-2.05 \pm 0.06$ & 
$-2.91 \pm 0.23$ & $-3.83 \pm 0.22$ & $+0.86 \pm 0.08$ & $+0.92 \pm 0.32$ \\
IRAS\,08339+6517 & $-0.63 \pm 0.08$ & $+0.16 \pm 0.07$ & $-1.45 \pm 0.09$ & 
$-2.15 \pm 0.14$ & $+0.29 \pm 0.13$ & $+1.61 \pm 0.11$ & $-2.44 \pm 0.19$ \\
Mrk\,33          & $-0.95 \pm 0.09$ & $-1.30 \pm 0.07$ & $-0.48 \pm 0.11$ & 
$-1.39 \pm 0.14$ & $-0.76 \pm 0.07$ & $-0.82 \pm 0.13$ & $-0.63 \pm 0.16$ \\
NGC\,1097        & $-0.37 \pm 0.13$ & $-0.75 \pm 0.12$ & $+0.77 \pm 0.12$ & 
$+0.91 \pm 0.17$ & $-0.65 \pm 0.15$ & $-1.51 \pm 0.17$ & $+1.56 \pm 0.23$ \\
NGC\,1313        & $-1.21 \pm 0.19$ & $-0.58 \pm 0.20$ & $-0.87 \pm 0.14$ & 
$-3.33 \pm 0.28$ & $-1.72 \pm 0.22$ & $+0.29 \pm 0.24$ & $-1.61 \pm 0.36$ \\
NGC\,1510        & $-1.25 \pm 0.08$ & $-0.62 \pm 0.07$ & $-0.92 \pm 0.08$ & 
$-1.54 \pm 0.17$ & $-1.74 \pm 0.14$ & $+0.30 \pm 0.10$ & $+0.21 \pm 0.22$ \\
NGC\,1705        & $-2.36 \pm 0.04$ & $-2.10 \pm 0.04$ & $-2.13 \pm 0.06$ & 
$-1.91 \pm 0.05$ & $-3.56 \pm 0.05$ & $+0.03 \pm 0.07$ & $+1.65 \pm 0.07$ \\
NGC\,1741        & $-0.16 \pm 0.13$ & $+1.15 \pm 0.11$ & $-0.64 \pm 0.11$ & 
$-0.71 \pm 0.14$ & $-2.58 \pm 0.18$ & $+1.79 \pm 0.16$ & $+1.87 \pm 0.23$ \\
NGC\,2782        & $-0.79 \pm 0.12$ & $-0.36 \pm 0.11$ & $-0.50 \pm 0.10$ & 
$+0.31 \pm 0.24$ & $-0.07 \pm 0.26$ & $+0.14 \pm 0.14$ & $+0.38 \pm 0.35$ \\
NGC\,3049        & $-1.23 \pm 0.12$ & $-0.65 \pm 0.08$ & $-0.43 \pm 0.12$ & 
$-2.20 \pm 0.27$ & $-0.94 \pm 0.17$ & $-0.22 \pm 0.14$ & $-1.27 \pm 0.32$ \\
NGC\,3353        & $-0.84 \pm 0.07$ & $-0.37 \pm 0.07$ & $-1.42 \pm 0.06$ & 
$-2.50 \pm 0.21$ & $-1.68 \pm 0.15$ & $+1.05 \pm 0.09$ & $-0.82 \pm 0.26$ \\
NGC\,3738        & $-1.47 \pm 0.07$ & $-0.94 \pm 0.06$ & $-1.11 \pm 0.05$ & 
$-0.97 \pm 0.14$ & $-2.06 \pm 0.12$ & $+0.17 \pm 0.08$ & $+1.09 \pm 0.18$ \\
NGC\,4214        & $-0.98 \pm 0.05$ & $-0.24 \pm 0.04$ & $-1.20 \pm 0.03$ & 
$-1.07 \pm 0.07$ & $-2.35 \pm 0.06$ & $+0.97 \pm 0.05$ & $+1.29 \pm 0.10$ \\
NGC\,4385        & $-0.16 \pm 0.12$ & $+0.10 \pm 0.12$ & $-0.88 \pm 0.09$ & 
$+0.06 \pm 0.14$ & $-2.88 \pm 0.12$ & $+0.98 \pm 0.15$ & $+2.94 \pm 0.18$ \\
NGC\,4449        & $-1.31 \pm 0.05$ & $-0.56 \pm 0.03$ & $-1.19 \pm 0.04$ & 
$-1.21 \pm 0.09$ & $-3.56 \pm 0.09$ & $+0.63 \pm 0.05$ & $+2.35 \pm 0.12$ \\
NGC\,4670        & $-1.23 \pm 0.06$ & $-0.97 \pm 0.05$ & $-1.45 \pm 0.05$ & 
$-0.21 \pm 0.09$ & $-1.77 \pm 0.07$ & $+0.49 \pm 0.07$ & $+1.56 \pm 0.12$ \\
NGC\,4861        & $-2.03 \pm 0.05$ & $-1.54 \pm 0.05$ & $-2.00 \pm 0.05$ & 
$-2.28 \pm 0.07$ & $-2.30 \pm 0.07$ & $+0.45 \pm 0.07$ & $+0.01 \pm 0.10$ \\
NGC\,6052        & $+0.39 \pm 0.18$ & $+2.38 \pm 0.15$ & $-0.91 \pm 0.13$ & 
$-0.53 \pm 0.18$ & $-1.47 \pm 0.14$ & $+3.29 \pm 0.20$ & $+0.94 \pm 0.23$ \\
NGC\,7496        & $-1.07 \pm 0.12$ & $-1.26 \pm 0.10$ & $+0.19 \pm 0.11$ & 
$-3.71 \pm 0.15$ & $-0.97 \pm 0.10$ & $-1.45 \pm 0.14$ & $-2.74 \pm 0.18$ \\
NGC\,7673        & $-0.96 \pm 0.09$ & $-0.50 \pm 0.09$ & $-0.63 \pm 0.08$ & 
$-2.29 \pm 0.16$ & $-2.05 \pm 0.11$ & $+0.13 \pm 0.12$ & $-0.24 \pm 0.20$ \\
NGC\,7714        & $-0.95 \pm 0.05$ & $-0.90 \pm 0.04$ & $-0.37 \pm 0.05$ & 
$-2.29 \pm 0.16$ & $-1.24 \pm 0.14$ & $-0.52 \pm 0.06$ & $-1.05 \pm 0.21$ \\
NGC\,7793        & $-0.69 \pm 0.13$ & $+0.43 \pm 0.13$ & $-1.17 \pm 0.10$ & 
$-4.42 \pm 0.19$ & $-2.13 \pm 0.18$ & $+1.60 \pm 0.16$ & $-2.29 \pm 0.26$ \\
UGC\,9560        & $-1.99 \pm 0.06$ & $-1.74 \pm 0.05$ & $-1.54 \pm 0.07$ & 
$-1.27 \pm 0.12$ & $-2.51 \pm 0.13$ & $-0.20 \pm 0.08$ & $+1.24 \pm 0.18$ \\
\noalign{\smallskip}
\hline
\end{tabular}
\end{table*}

Here we test the reliability of our method of investigating the presence of 
dust with an extinction law exhibiting the 2175\,\AA{} bump, that is based on 
a new parametric approach. We do so by investigating the slope parameters 
defined in Sect.~\ref{technique} of our IUE sample of nearby starbursts (see 
Sect.~\ref{IUE}) with respect to models. Table~\ref{tab_IUE_gamma} lists the 
parameters measured by us for this reference sample. Composite stellar 
populations models are calculated as in \cite{MAR05} for a standard Salpeter 
IMF, a continuous star-formation activity along 50 or 500 Myr, and a 
metallicity equal to $0.5$\,Z$_{\odot}$ or $1.0$\,Z$_{\odot}$. Furthermore, 
the properties of dust attenuation are described through the two-phase, 
clumpy SHELL dust/stars configuration of \cite{WIT00} for SMC-like or MW-like 
dust, where the average opacity of the dust shell is a free parameter.  

The combined slope parameter $\gamma_{34}$ is most suitable to study the 
presence of the 2175\,\AA{} feature. This is illustrated by the plot of 
$\gamma_{34}$ versus $\beta$ in Fig.~\ref{fig_gambet_test} which shows 
that models with different composite stellar populations and SMC-type dust,
lacking the 2175\,\AA{} feature, exhibit $\gamma_{34} \sim 1$, whatever 
$\tau_V$ and $\beta$, respectively. Conversely, $\gamma_{34}$ rapidly 
decreases below zero when $\beta$ increases for models where UV-bump producing 
MW-type dust is present (see Fig.~\ref{fig_extinct}). Now, the 24 local 
starbursts of the IUE sample are distributed in a region delimited by 
$-2.5 < \beta < 0.5$ and $-3 < \gamma_{34} < 3$. The bulk of these galaxies 
populates a region of the $\beta - \gamma_{34}$ plane that is confined by the 
model curves, for $\beta > -3$. Our interpretation is that these nearby 
starbursts possibly contain relatively small amount of dust with an 
extinction law exhibiting only a modest UV bump, if any. Moreover, analysing 
the distributions of the other slope parameters (see Fig.~\ref{fig_gambet}) we 
find that the far-UV uprise of the extinction law is intermediate between 
those of the SMC and MW. These are the same conclusions of Gordon et al. 
(\cite{GOR97}) and \cite{WIT00} for the sample of 30 nearby starbursts with 
IUE data selected by Calzetti et al. (\cite{CAL94}). Hence, our method of 
analysis is proved to be reliable.

\section{Results}\label{results}

\begin{table*}
\caption[]{UV continuum slope parameters of our sample of 34 FDF galaxies at 
$2 < z < 2.5$ (see Table~\ref{tab_FDF_basic})}
\label{tab_FDF_gamma}
\centering
\begin{tabular}{c c c c c c c c}
\hline\hline
\noalign{\smallskip}
No. & $\beta$ & $\gamma_1$ & $\gamma_2$ & 
$\gamma_3$ & $\gamma_4$ & $\gamma_{12}$ & $\gamma_{34}$ \\
\noalign{\smallskip}
\hline
\noalign{\smallskip}
1208 & $-0.25 \pm 0.11$ & $-0.72 \pm 0.12$ & $+0.03 \pm 0.08$ & 
$-1.21 \pm 0.09$ & $-1.28 \pm 0.37$ & $-0.75 \pm 0.15$ & $+0.06 \pm 0.39$ \\
1691 & $-0.29 \pm 0.11$ & $-0.28 \pm 0.13$ & $-0.53 \pm 0.14$ & 
$-1.02 \pm 0.16$ & $-0.42 \pm 0.20$ & $+0.25 \pm 0.19$ & $-0.60 \pm 0.23$ \\
1744 & $-0.65 \pm 0.13$ & $+0.68 \pm 0.14$ & $-0.52 \pm 0.14$ & 
$-1.87 \pm 0.19$ & $-4.30 \pm 0.38$ & $+1.21 \pm 0.19$ & $+2.42 \pm 0.41$ \\
1991 & $-1.44 \pm 0.23$ & $-0.76 \pm 0.52$ & $-2.24 \pm 0.23$ & 
$-2.43 \pm 0.24$ & $-3.72 \pm 0.79$ & $+1.48 \pm 0.49$ & $+1.29 \pm 0.79$ \\
2274 & $-0.74 \pm 0.07$ & $-0.17 \pm 0.05$ & $-0.98 \pm 0.05$ & 
$-1.26 \pm 0.14$ & $-1.68 \pm 0.43$ & $+0.81 \pm 0.07$ & $+0.42 \pm 0.40$ \\
2418 & $-0.21 \pm 0.09$ & $-0.29 \pm 0.08$ & $-0.42 \pm 0.13$ & 
$-1.60 \pm 0.10$ & $+0.35 \pm 0.19$ & $+0.13 \pm 0.16$ & $-1.94 \pm 0.22$ \\
2495 & $-0.22 \pm 0.06$ & $-0.03 \pm 0.07$ & $-0.56 \pm 0.07$ & 
$-2.05 \pm 0.07$ & $+0.46 \pm 0.69$ & $+0.53 \pm 0.10$ & $-2.51 \pm 0.69$ \\
2636 & $+0.93\pm  0.14$ & $+2.37 \pm 0.15$ & $+0.38 \pm 0.09$ & 
$-2.83 \pm 0.15$ & $+1.67 \pm 0.63$ & $+1.98 \pm 0.17$ & $-4.49 \pm 0.71$ \\
3005 & $-0.14 \pm 0.09$ & $+0.16 \pm 0.07$ & $-0.18 \pm 0.09$ & 
$-2.34 \pm 0.10$ & $-0.65 \pm 0.27$ & $+0.33 \pm 0.11$ & $-1.69 \pm 0.29$ \\
3163 & $-0.19 \pm 0.08$ & $+0.44 \pm 0.07$ & $-0.74 \pm 0.07$ & 
$-2.34 \pm 0.09$ & $-0.07 \pm 0.82$ & $+1.19 \pm 0.10$ & $-2.27 \pm 0.82$ \\
3300 & $-1.36 \pm 0.08$ & $-1.28 \pm 0.08$ & $-1.43 \pm 0.11$ & 
$-2.08 \pm 0.15$ & $-3.27 \pm 0.66$ & $+0.15 \pm 0.14$ & $+1.19 \pm 0.64$ \\
3374 & $-1.63 \pm 0.06$ & $-1.48 \pm 0.07$ & $-1.56 \pm 0.07$ & 
$-1.44 \pm 0.11$ & $-1.23 \pm 0.27$ & $+0.08 \pm 0.10$ & $-0.21 \pm 0.29$ \\
3688 & $-0.33 \pm 0.11$ & $+1.19 \pm 0.09$ & $-1.88 \pm 0.10$ & 
$-3.54 \pm 0.16$ & $-2.96 \pm 0.59$ & $+3.07 \pm 0.13$ & $-0.58 \pm 0.61$ \\
3810 & $-0.33 \pm 0.06$ & $-0.03 \pm 0.04$ & $-0.70 \pm 0.09$ & 
$-2.94 \pm 0.07$ & $+0.30 \pm 0.24$ & $+0.67 \pm 0.09$ & $-3.24 \pm 0.25$ \\
3874 & $+0.01 \pm 0.09$ & $+0.94 \pm 0.08$ & $-1.07 \pm 0.09$ & 
$-4.16 \pm 0.11$ & $-1.43 \pm 0.99$ & $+2.01 \pm 0.12$ & $-2.73 \pm 1.00$ \\
3875 & $-1.52 \pm 0.11$ & $-1.33 \pm 0.10$ & $-0.61 \pm 0.12$ & 
$-2.47 \pm 0.12$ & $-1.34 \pm 0.36$ & $-0.72 \pm 0.16$ & $-1.14 \pm 0.35$ \\
3958 & $-0.65 \pm 0.10$ & $-0.18 \pm 0.11$ & $-1.75 \pm 0.10$ & 
$-2.03 \pm 0.13$ & $-3.13 \pm 0.27$ & $+1.56 \pm 0.15$ & $+1.09 \pm 0.30$ \\
4795 & $-0.78 \pm 0.08$ & $+0.51 \pm 0.07$ & $-1.04 \pm 0.06$ & 
$-1.80 \pm 0.08$ & $-1.86 \pm 0.15$ & $+1.55 \pm 0.09$ & $+0.05 \pm 0.18$ \\
4871 & $-0.09 \pm 0.09$ & $+0.27 \pm 0.09$ & $-0.61 \pm 0.09$ & 
$-2.75 \pm 0.13$ & $+1.15 \pm 0.48$ & $+0.89 \pm 0.12$ & $-3.89 \pm 0.50$ \\
4996 & $+0.03 \pm 0.07$ & $+1.00 \pm 0.05$ & $-0.50 \pm 0.06$ & 
$-1.94 \pm 0.09$ & $-1.19 \pm 0.09$ & $+1.50 \pm 0.08$ & $-0.74 \pm 0.13$ \\
5058 & $-1.03 \pm 0.07$ & $-0.44 \pm 0.05$ & $-0.90 \pm 0.07$ & 
$-1.55 \pm 0.05$ & $-2.39 \pm 0.10$ & $+0.45 \pm 0.08$ & $+0.84 \pm 0.11$ \\
5135 & $-0.37 \pm 0.13$ & $-0.54 \pm 0.11$ & $-1.51 \pm 0.15$ & 
$-2.75 \pm 0.18$ & $-2.79 \pm 0.32$ & $+0.97 \pm 0.19$ & $+0.04 \pm 0.34$ \\
5165 & $-1.06 \pm 0.09$ & $-0.83 \pm 0.09$ & $-1.48 \pm 0.13$ & 
$-3.03 \pm 0.17$ & $-2.90 \pm 0.30$ & $+0.65 \pm 0.16$ & $-0.14 \pm 0.37$ \\
5190 & $-1.91 \pm 0.11$ & $-0.79 \pm 0.07$ & $-1.56 \pm 0.13$ & 
$-2.69 \pm 0.24$ & $-1.17 \pm 0.39$ & $+0.77 \pm 0.15$ & $-1.53 \pm 0.38$ \\
5227 & $+0.02 \pm 0.15$ & $+1.40 \pm 0.12$ & $-1.32 \pm 0.13$ & 
$-4.85 \pm 0.18$ & $-0.02 \pm 0.29$ & $+2.72 \pm 0.18$ & $-4.83 \pm 0.36$ \\
6024 & $-0.99 \pm 0.06$ & $-0.64 \pm 0.06$ & $-0.84 \pm 0.06$ & 
$-2.61 \pm 0.07$ & $-1.33 \pm 0.35$ & $+0.20 \pm 0.08$ & $-1.28 \pm 0.36$ \\
6372 & $-0.79 \pm 0.07$ & $-0.69 \pm 0.06$ & $-0.69 \pm 0.10$ & 
$-1.25 \pm 0.08$ & $-1.39 \pm 0.18$ & $-0.01 \pm 0.13$ & $+0.14 \pm 0.19$ \\
6407 & $+1.02 \pm 0.16$ & $+2.69 \pm 0.17$ & $-0.40 \pm 0.12$ & 
$-0.59 \pm 0.16$ & $-2.21 \pm 0.28$ & $+3.09 \pm 0.21$ & $+1.62 \pm 0.32$ \\
6934 & $-0.26 \pm 0.09$ & $+0.54 \pm 0.08$ & $-0.20 \pm 0.09$ & 
$-3.25 \pm 0.14$ & $-3.18 \pm 0.91$ & $+0.75 \pm 0.12$ & $-0.09 \pm 0.92$ \\
6947 & $-0.10 \pm 0.08$ & $+0.65 \pm 0.07$ & $-0.84 \pm 0.12$ & 
$-3.45 \pm 0.12$ & $+0.00 \pm 0.15$ & $+1.49 \pm 0.13$ & $-3.46 \pm 0.19$ \\
7029 & $-0.68 \pm 0.07$ & $-0.35 \pm 0.06$ & $-0.79 \pm 0.18$ & 
$-2.99 \pm 0.14$ & $-0.72 \pm 0.23$ & $+0.43 \pm 0.18$ & $-2.26 \pm 0.27$ \\
7078 & $-0.41 \pm 0.19$ & $-0.49 \pm 0.53$ & $-0.45 \pm 0.15$ & 
$-1.81 \pm 0.42$ & $-4.58 \pm 0.77$ & $-0.04 \pm 0.54$ & $+2.70 \pm 0.98$ \\
7307 & $-0.56 \pm 0.11$ & $-0.19 \pm 0.13$ & $-1.55 \pm 0.10$ & 
$-0.34 \pm 0.17$ & $-2.48 \pm 0.85$ & $+1.35 \pm 0.17$ & $+2.19 \pm 0.86$ \\
7342 & $-0.08 \pm 0.16$ & $-0.42 \pm 0.15$ & $+0.28 \pm 0.20$ & 
$-4.05 \pm 0.20$ & $+0.09 \pm 0.78$ & $-0.70 \pm 0.28$ & $-4.14 \pm 0.79$ \\
\noalign{\smallskip}
\hline
\end{tabular}
\end{table*}

\begin{table*}
\caption[]{Average UV continuum slope parameters of the local IUE sample and
different $2 < z < 2.5$ FDF samples. The slope parameters were averaged 
unweighted. The uncertainties given are mean errors and were derived from the 
variance of the sample without considering the individual errors.}
\label{tab_gamma_mean}
\centering
\begin{tabular}{c c c c c c c}
\hline\hline
\noalign{\smallskip}
Par. & IUE & FDF & FDF            & FDF                  &
FDF                  & FDF                  \\
     & all & all & $\beta < -0.4$ & $-0.4 < \beta < 0.1$ & 
$-0.4 < \beta < 0.1$ & $-0.4 < \beta < 0.1$ \\
     &     &     &                &                      &                
$\gamma_{34} > -2$   & $\gamma_{34} < -2$   \\ 
\noalign{\smallskip}
\hline
\noalign{\smallskip}
$N$           & 24               & 34               & 16               &
16               & 8                & 8                \\
$\beta$       & $-1.05 \pm 0.13$ & $-0.50 \pm 0.11$ & $-1.01 \pm 0.11$ &
$-0.18 \pm 0.03$ & $-0.23 \pm 0.04$ & $-0.12 \pm 0.04$ \\
$\gamma_1$    & $-0.51 \pm 0.16$ & $+0.03 \pm 0.16$ & $-0.53 \pm 0.15$ &
$+0.27 \pm 0.16$ & $+0.13 \pm 0.25$ & $+0.40 \pm 0.21$ \\
$\gamma_2$    & $-0.98 \pm 0.14$ & $-0.86 \pm 0.11$ & $-1.15 \pm 0.13$ &
$-0.67 \pm 0.14$ & $-0.65 \pm 0.24$ & $-0.69 \pm 0.17$ \\
$\gamma_3$    & $-1.63 \pm 0.26$ & $-2.33 \pm 0.17$ & $-1.98 \pm 0.18$ &
$-2.76 \pm 0.27$ & $-2.21 \pm 0.33$ & $-3.32 \pm 0.34$ \\ 
$\gamma_4$    & $-1.90 \pm 0.22$ & $-1.46 \pm 0.26$ & $-2.34 \pm 0.30$ &
$-0.73 \pm 0.33$ & $-1.51 \pm 0.46$ & $+0.06 \pm 0.26$ \\
$\gamma_{12}$ & $+0.46 \pm 0.21$ & $+0.88 \pm 0.17$ & $+0.62 \pm 0.17$ & 
$+0.94 \pm 0.27$ & $+0.78 \pm 0.40$ & $+1.10 \pm 0.36$ \\
$\gamma_{34}$ & $+0.27 \pm 0.32$ & $-0.87 \pm 0.35$ & $+0.36 \pm 0.36$ &
$-2.04 \pm 0.40$ & $-0.69 \pm 0.27$ & $-3.38 \pm 0.31$ \\
\noalign{\smallskip}
\hline
\end{tabular}
\end{table*}

\begin{figure*}
\centering
\mbox{
\includegraphics[width=7.68cm,clip=true]{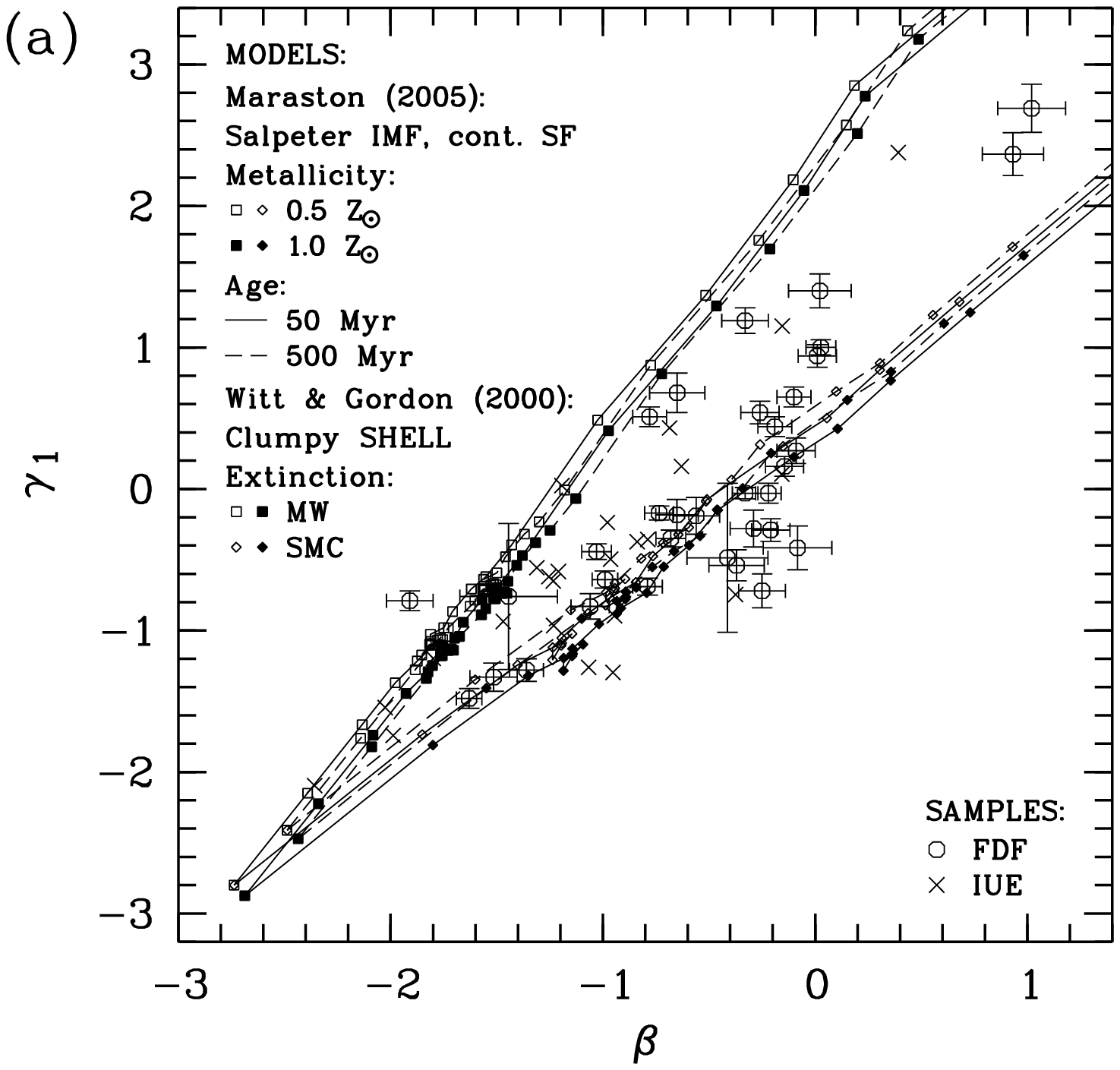}\hspace{0.64cm}
\includegraphics[width=7.68cm,clip=true]{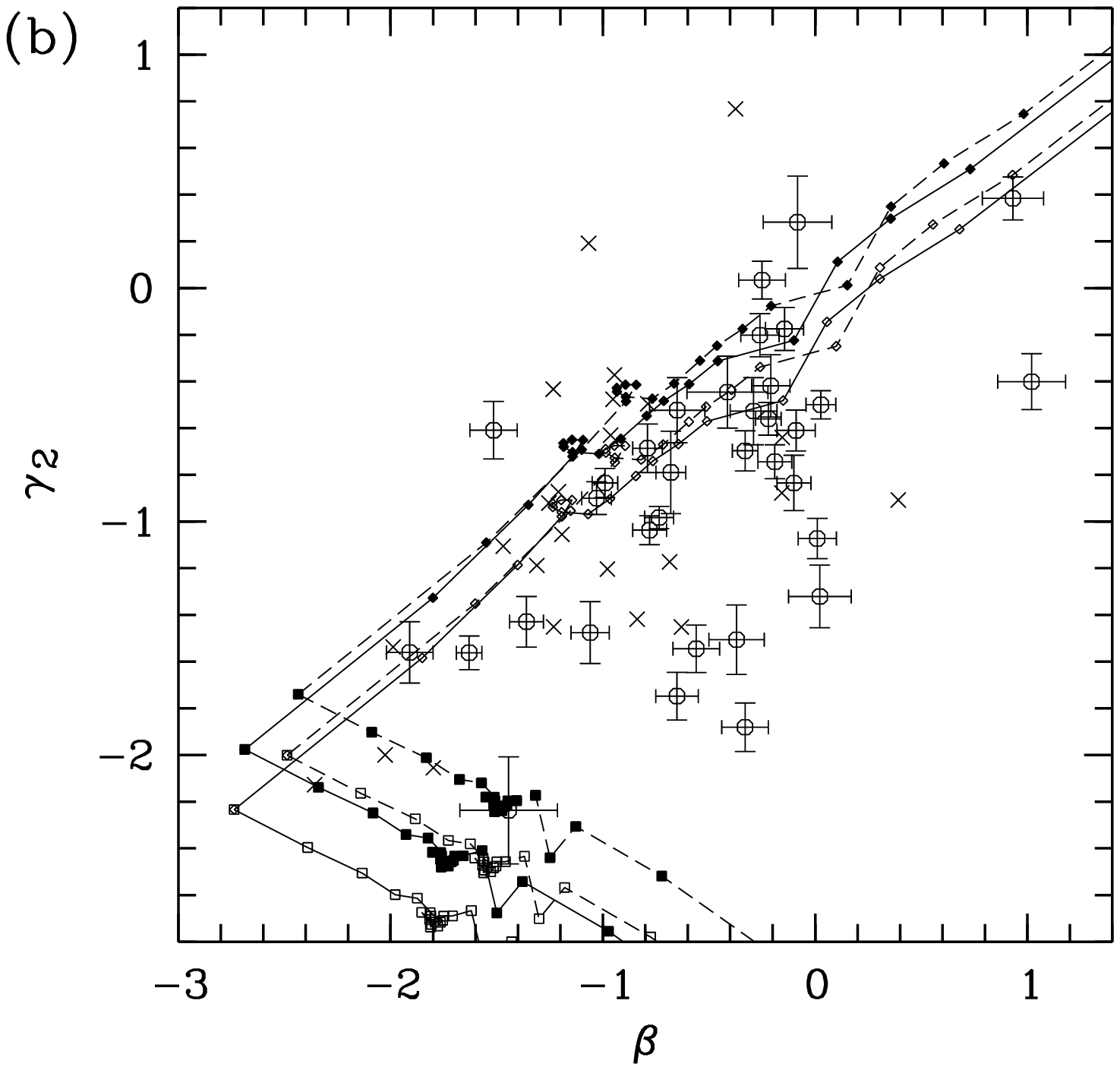}}\\
\mbox{
\includegraphics[width=7.68cm,clip=true]{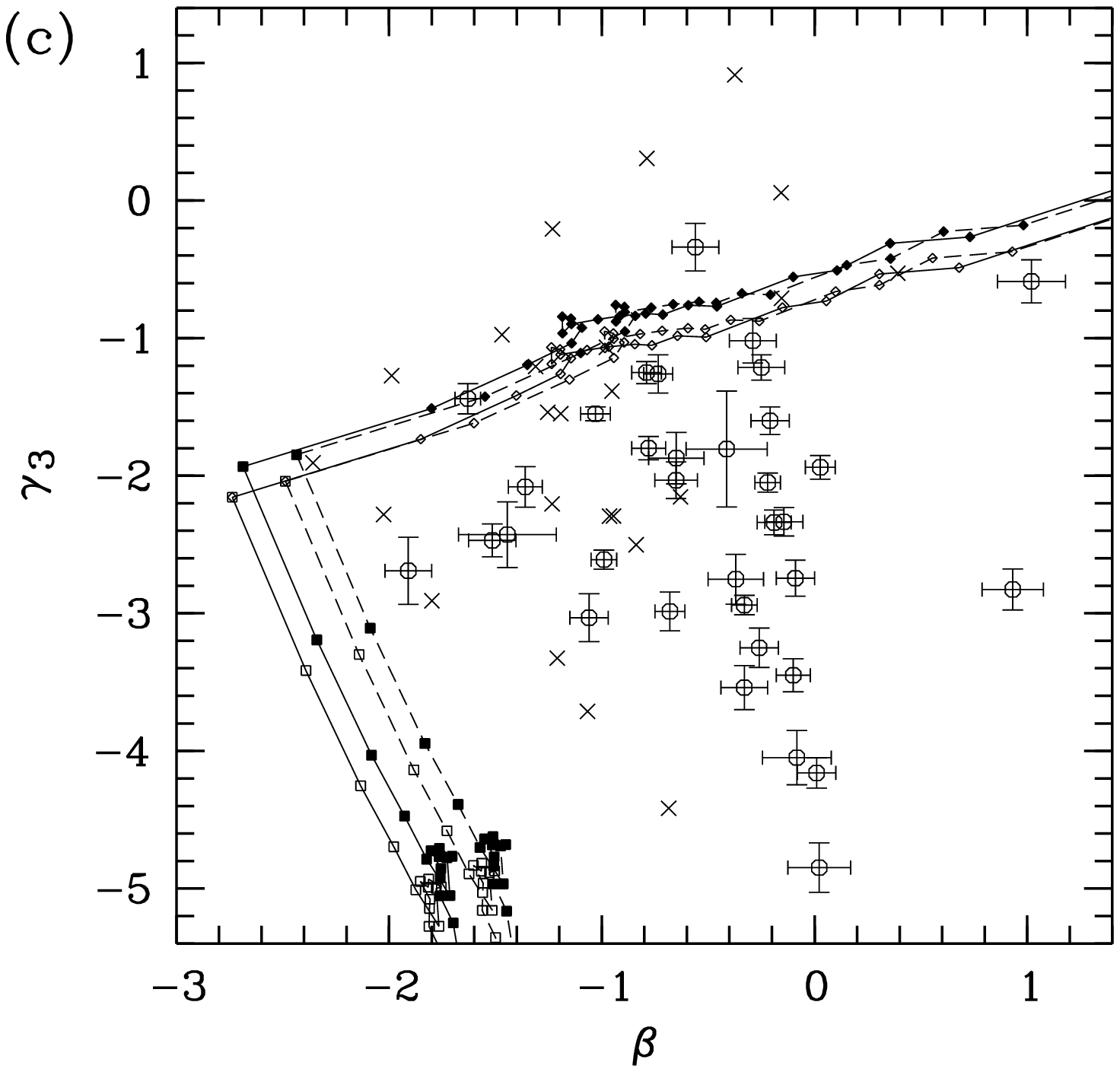}\hspace{0.64cm}
\includegraphics[width=7.68cm,clip=true]{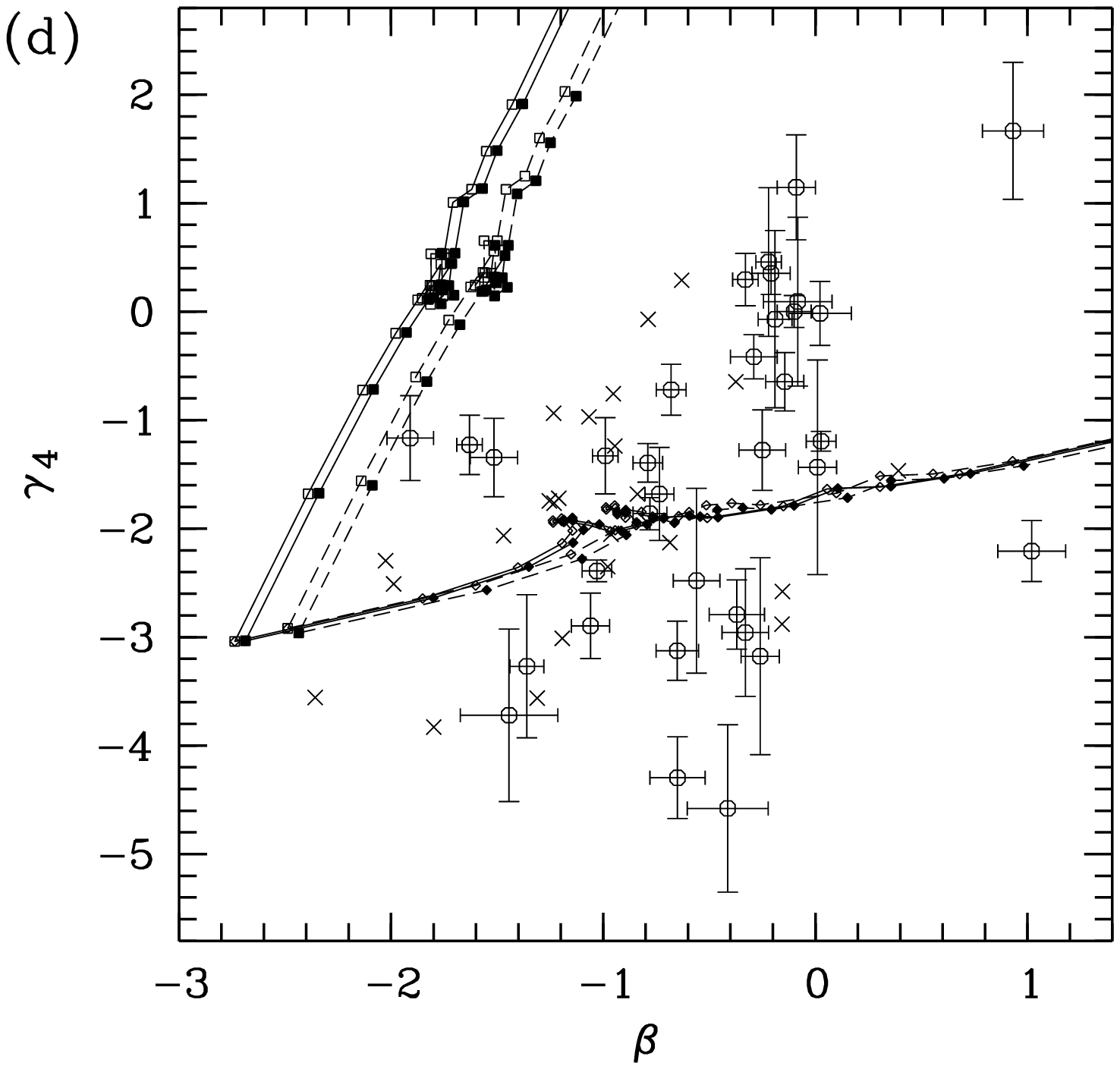}}\\
\mbox{
\includegraphics[width=7.68cm,clip=true]{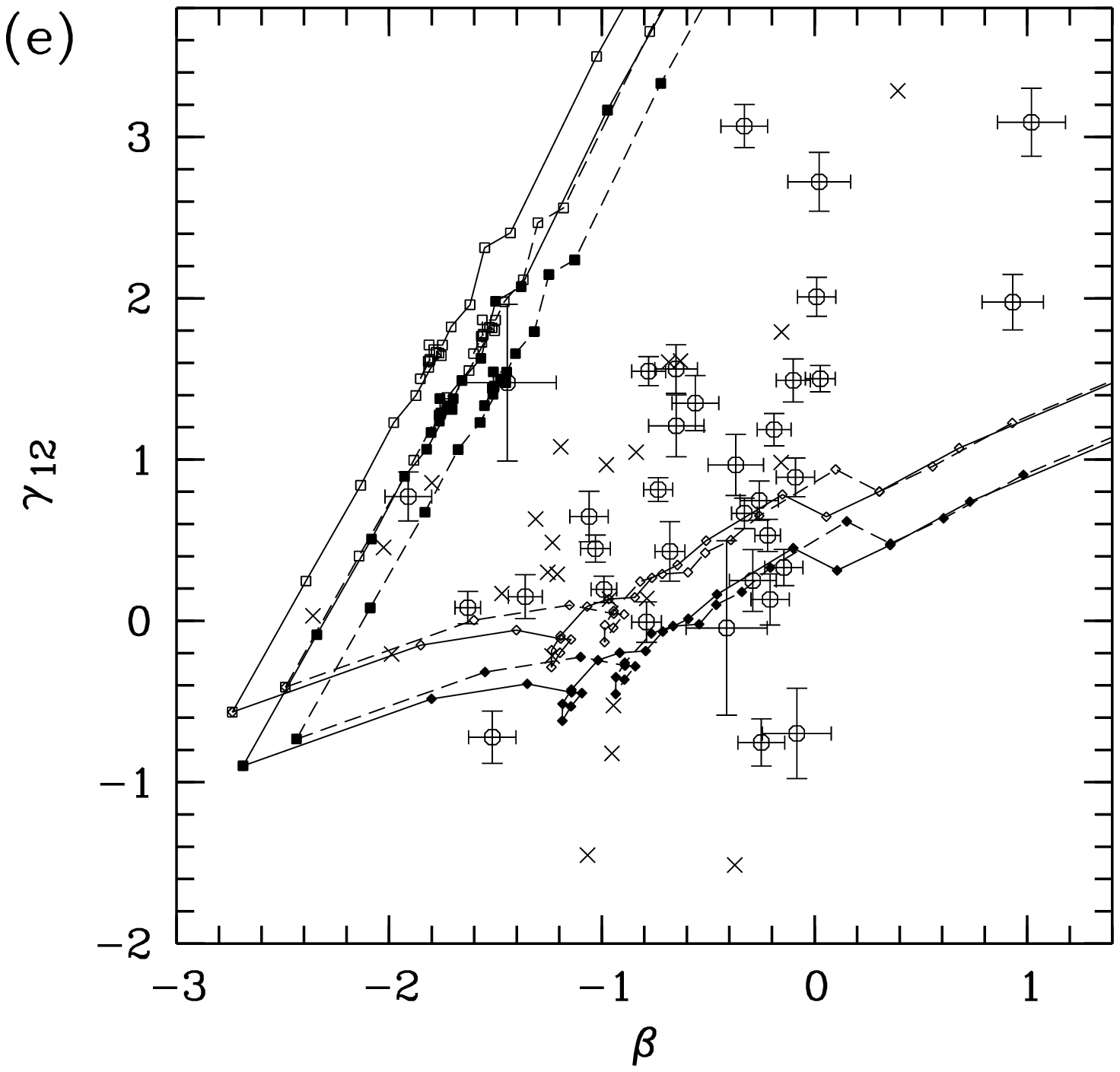}\hspace{0.64cm}
\includegraphics[width=7.68cm,clip=true]{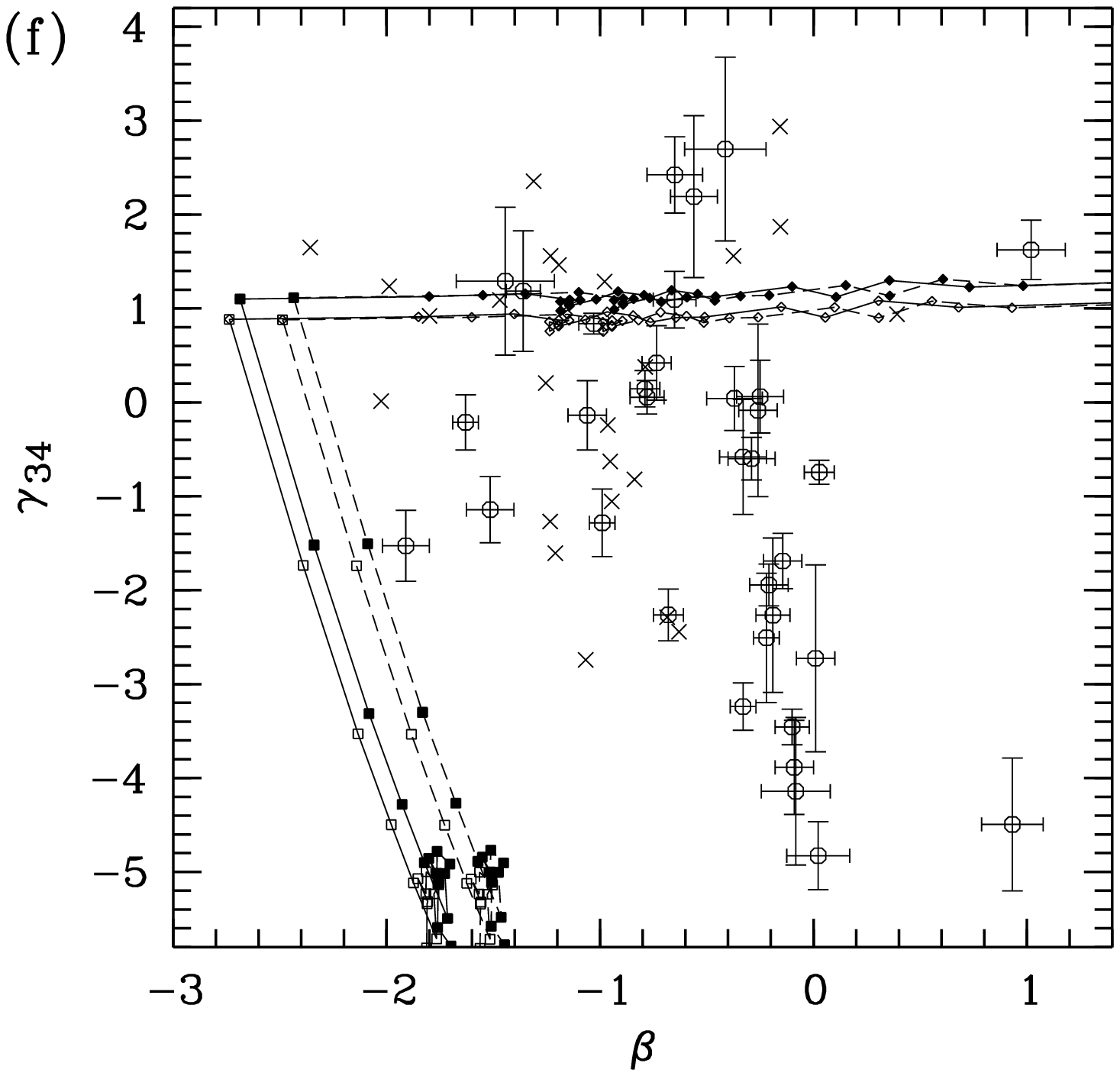}}
\caption[]{The slope parameters $\gamma_1$ to $\gamma_4$, 
$\gamma_{12} = \gamma_1 - \gamma_2$, and  
$\gamma_{34} = \gamma_3 - \gamma_4$ versus $\beta$ for the FDF sample of 
$2 < z < 2.5$ galaxies (open circles with error bars) and the IUE sample of 
local starburst galaxies (crosses). The spectroscopic data are compared to 
combined \cite{MAR05} stellar population synthesis and \cite{WIT00} radiative
transfer models. Details of the models shown are indicated in the legend of 
subfigure (a). The free parameter of the curves is the average radial visual 
optical depth $\tau_V$ (see Fig.~\ref{fig_beta_A1500}).}
\label{fig_gambet} 
\end{figure*}

Table~\ref{tab_FDF_gamma} lists the values of the UV continuum slopes $\beta$
and $\gamma_1$ to $\gamma_4$, together with the parameters $\gamma_{12}$ and 
$\gamma_{34}$ measured for the FDF sample of 34 UV-luminous galaxies at 
$2 < z < 2.5$ (see Sect.~\ref{FDF}). Hereafter, we will investigate the 
distribution of these galaxies in diagnostic diagrams defined by $\beta$ and 
one of the six $\gamma$ parameters in comparison with that of the 24 local 
starbursts of the IUE sample (see Sect.~\ref{IUE} and 
Table~\ref{tab_IUE_gamma}) and of models combining stellar population 
synthesis and dust attenuation as in Sect.~\ref{test}.

Meurer et al. (\cite{MEU99}) proposed the $\beta$ parameter as a measure of 
dust attenuation at UV wavelengths for nearby and distant starburst galaxies. 
More recently, Leitherer et al. (\cite{LEI02}) determine 
\begin{equation}\label{eq_A1500}
A_{1500} [{\rm mag}] = 2.19\,(\beta - \beta_0),\ \beta_0 = -2.5
\end{equation}
as the relation between $\beta$ and the attenuation at 1500\,\AA{} for nearby 
starburst galaxies. However, radiative transfer modelling (\cite{WIT00}) 
demonstrates that the ``standard'' $\beta$ parameter is a model-dependent 
measure of dust attenuation at UV wavelengths. In particular, 
Fig.~\ref{fig_beta_A1500} illustrates the relation between $\beta$ and 
$A_{1500}$ for the \cite{WIT00} models discussed in Sect.~\ref{radtransfer}. 
Furthermore, it shows that $\beta$ is not a linear measure of dust 
attenuation at UV wavelengths along the whole range potentially available for 
$A_{1500}$ for a sample of objects, even when dust attenuation can be 
described in the same way for all these objects. With these caveats in mind, 
we determine the mean value of $\beta$ to be 
$\langle\beta\rangle = -1.05 \pm 0.13$ for the IUE sample and 
$\langle\beta\rangle = -0.50 \pm 0.11$ for the FDF sample (see 
Table~\ref{tab_gamma_mean}). If the stellar populations, the dust/stars 
configuration, and the structure of the dusty ISM are the same for the 
galaxies of both samples (see Sect.~\ref{models}), then 
Fig.~\ref{fig_beta_A1500} implies that the mean attenuation at UV wavelengths 
is larger for the FDF galaxies (see also Noll et al. \cite{NOL04}). In this 
case, we probably compare $z \sim 2$ ultraluminous galaxies (i.e. with 
$L_{\rm UV} \gtrsim 10^{12}$\,L$_{\odot}$) to fainter (by about two orders of 
magnitude) local starburst galaxies (see Heckman et al. \cite{HEC98}; 
Leitherer et al. \cite{LEI02}), given the gap in observed UV luminosities 
(see Fig.~\ref{fig_luvz}).

Fig.~\ref{fig_gambet} shows the distribution of the FDF and IUE sample 
galaxies and the loci of the models in several diagrams where the individual 
$\gamma$ parameters are plotted versus $\beta$. Open circles and crosses 
reproduce, respectively, the $z \sim 2$ UV-luminous FDF galaxies and the 
local IUE starbursts. For both samples, the measured parameters have similar
uncertainties, though for the FDF sample the most uncertain one is $\gamma_4$ 
due to the bright night sky in the corresponding wavelength range. For the 
IUE sample the uncertainties of $\gamma_3$ are particularly high due to the 
low quality of the combined spectra in the overlapping region of the IUE 
short and long wavelength channels (see Sect.~\ref{IUE}). Therefore, only 
error bars for the FDF galaxies are drawn to ease the reading of 
Fig.~\ref{fig_gambet}. Different models are reproduced synthetically as 
illustrated in Fig.~\ref{fig_gambet}(a). We discuss the results obtained from
the analysis of Fig.~\ref{fig_gambet} in the next two sections.

\subsection{The far-UV slope of the extinction law}\label{curvature}

Panels (a), (b), and (e) of Fig.~\ref{fig_gambet} show, respectively,
$\gamma_1$, $\gamma_2$, and $\gamma_{12}$ versus $\beta$ both for the 
observed spectra and the synthetic ones. As the models show, the distribution 
in these three diagrams characterises the far-UV slope of the extinction law. 
Not surprisingly, both the models and the observed galaxies are distributed 
almost along the one-to-one relation in the $\beta - \gamma_1$ plane. This is 
due to the large overlap between the wavelength windows where the two slopes 
are defined (see Fig.~\ref{fig_demogamma}). Models with different extinction 
curves split up much better in the $\beta - \gamma_2$ and 
$\beta - \gamma_{12}$ planes. In fact, $\gamma_2$ and $\beta$ are defined in 
wavelength regions with only a partial overlap.

Interestingly, most of the FDF and IUE galaxies are distributed in the region
delimited by the models with MW or SMC dust in each diagram. However, the 
baricenter of the galaxy distribution is close to the locus of the models 
with SMC extinction law, despite the scatter of the distribution is large. We 
conclude that the mixture of dust grains present in the UV-luminous FDF 
galaxies at $z \sim 2$ produces an extinction curve with a far-UV slope that 
is intermediate between those of the MW and SMC extinction curves, if not 
close to the far-UV slope of the SMC extinction curve.

This result is robust against differences in the stellar populations induced
by metallicity, age, and star-formation history of the real galaxies, for the 
reasons discussed in Sect.~\ref{sensitivity}. Small differences do exist 
between the distributions of models and galaxies reproduced in panels (a), 
(b), and (e) of Fig.~\ref{fig_gambet} however. For instance, $\gamma_1$ is 
sensitive to the maximum age of the stellar populations (i.e. the age of the 
model), while $\gamma_2$ is sensitive to the metallicity of the stellar 
populations (see Table~\ref{tab_dgamma_sps}). Therefore, individual galaxies 
will shift slightly with respect to the loci of models with different dust 
types in the $\beta - \gamma_1$, $\beta - \gamma_2$, and 
$\beta - \gamma_{12}$ planes according to the true characteristic age and 
metallicity of their stellar populations.

Furthermore, we note that 26 of 34 UV-luminous FDF galaxies at $z \sim 2$
exhibit $\beta > -0.8$, which can be achieved by the two-phase, clumpy 
SHELL models with SMC dust for $\tau_V > 3 - 5$, depending on the properties 
of the stellar population. In this case the dusty clumps are optically thick 
at UV wavelengths, and the UV photons escaping from the system do so after 
scattering through the diffuse interclump medium that is still optically thin 
(see Sect.~\ref{radtransfer}). The values of the optical depths of a typical 
clump and of the diffuse interclump medium depend on the filling factor and 
density contrast assumed, once $\tau_V$ is fixed (see Witt \& Gordon 
\cite{WIT96}, \cite{WIT00}). Nevertheless, multiple scattering does play a 
role in the escape of the UV photons from the reddest, dustiest galaxies of 
the FDF sample under study.

\subsection{Evidence for the presence of the UV bump}\label{bump}

Panels (c), (d), and (f) of Fig.~\ref{fig_gambet} show, respectively,
$\gamma_3$, $\gamma_4$, and $\gamma_{34}$ versus $\beta$ both for
the observed spectra and the synthetic ones. The parameters $\gamma_3$
and $\gamma_4$ are defined in wavelength windows corresponding to,
respectively, the blue and red wings of the 2175\,\AA{} feature
present in the average MW and LMC extinction curves. As the models show,
the parameters $\gamma_3$ and $\gamma_{34}$ are the most sensitive
to the presence of the carriers of the UV bump as determined from
an observed spectrum. Furthermore, we recall that $\gamma_3$ and $\gamma_4$
are affected by relatively large errors for, respectively, the IUE and FDF 
samples.

Now, Fig.~\ref{fig_gambet}(c) and (f) clearly show that the $z \sim 2$
UV-luminous FDF galaxies populate a region delimited by models with either
a MW- or an SMC-type extinction curve. Furthermore, Fig.~\ref{fig_gambet}(f)
suggests that $\gamma_{34}$ becomes more negative when $\beta$ increases.
This trend means that the 2175\,\AA{} feature characterises the redder,
i.e. more attenuated and dustier, FDF galaxies. Conversely, the IUE nearby
starbursts are almost equally distributed across the locus of models with 
SMC-type dust.

These findings are consistent with those illustrated in the previous section,
based on $\gamma$ parameters defined in distinct windows corresponding
to the short-wavelength region of the UV spectral domain. Hence, we conclude
that the redder, UV-luminous FDF galaxies at $2 < z < 2.5$ contain a mixture
of dust grains producing an extinction curve that has both a far-UV slope and 
a strength of the UV bump that are intermediate between those of the SMC- and 
MW-type extinction curves.

\subsection{Different dust properties for different galaxy subsamples}
\label{subsamples}

\begin{figure}
\centering 
\includegraphics[width=8.8cm,clip=true]{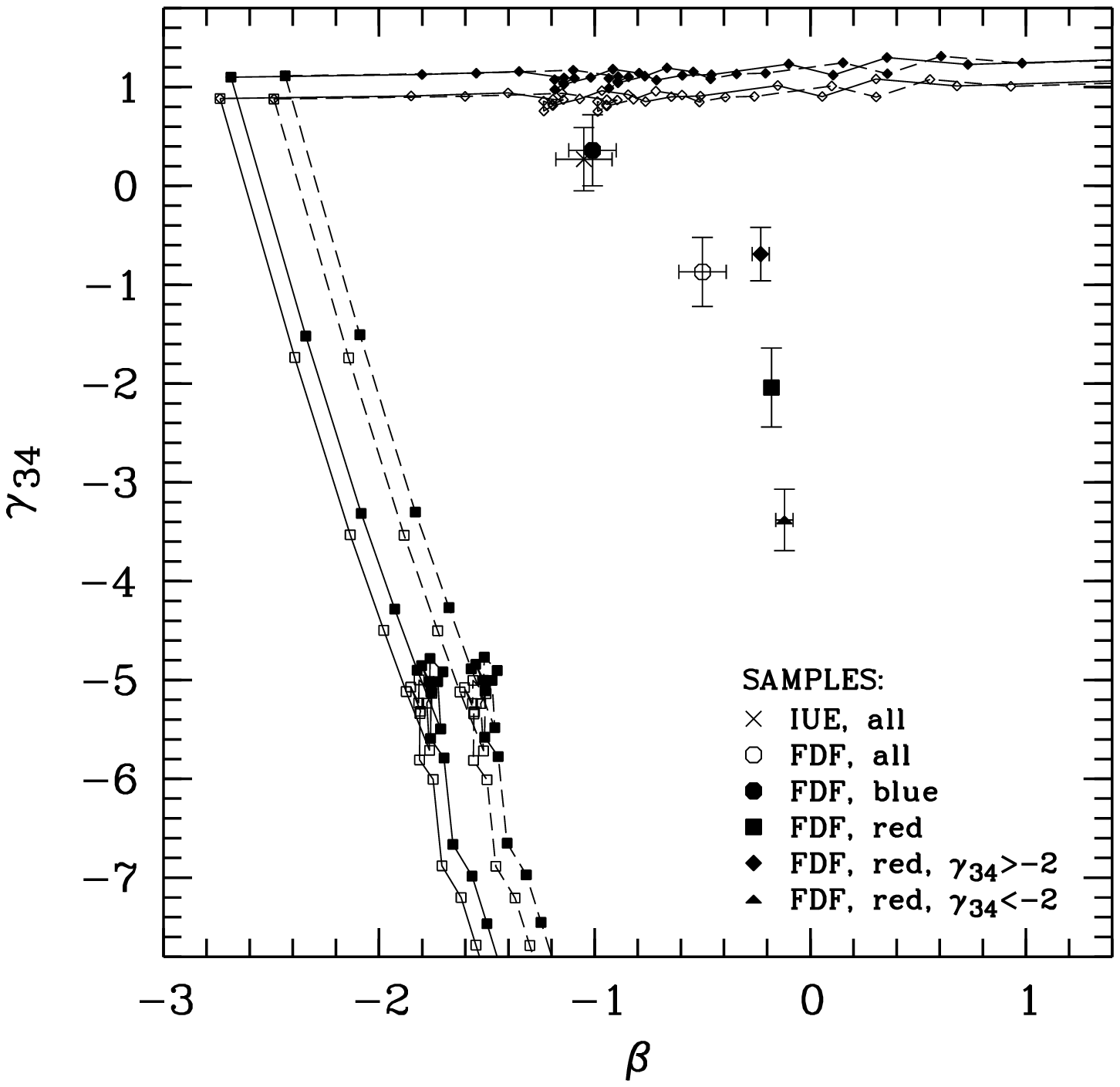}
\caption[]{The UV-bump indicator $\gamma_{34}$ as a function of the 
reddening measure $\beta$ for our local galaxy sample (cross) and different 
subsamples of $2 < z < 2.5$ FDF galaxies. The complete FDF sample of 34 
galaxies (open circle) was divided into a blue ($\beta < -0.4$; 16 objects; 
filled circle) and a red subsample ($-0.4 < \beta < 0.1$; 16 objects; filled 
square) excluding two very red objects with $\beta \sim 1$. Moreover, the red 
sample was further fragmented into two subsamples of eight objects each 
differing in $\gamma_{34}$. The filled lozenge and the filled triangle mark 
galaxies with $\gamma_{34} > -2$ and $\gamma_{34} < -2$, respectively. The 
$1\,\sigma$ error bars indicate mean errors. The diagram also shows combined 
stellar population synthesis and radiative transfer models as a function of 
the optical depth of the dust. Details of the models shown are indicated in 
the legend of Fig.~\ref{fig_gambet}(a).}
\label{fig_gam34bet_mean}
\end{figure}

\begin{figure}
\centering 
\includegraphics[width=8.8cm,clip=true]{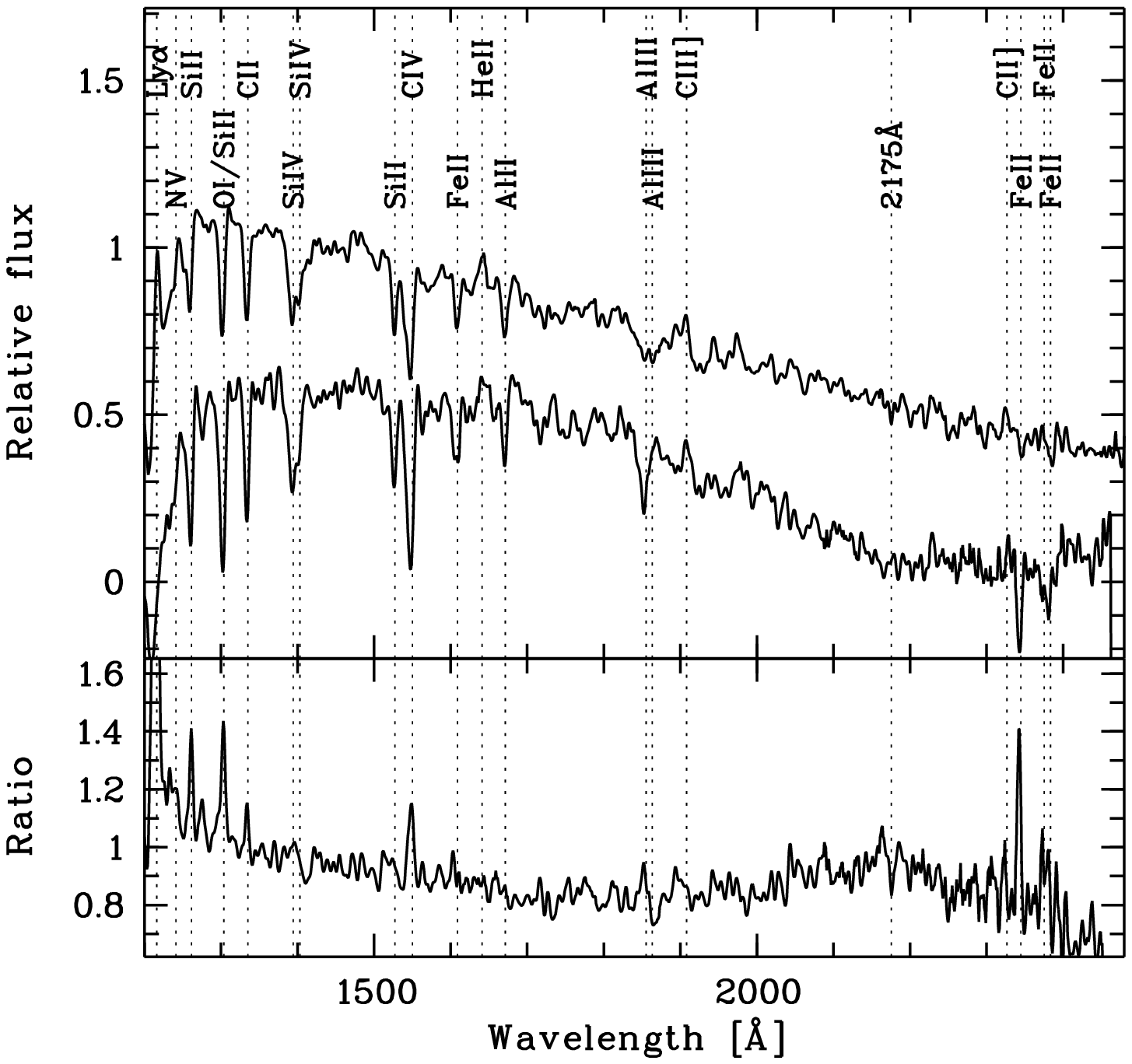}
\caption[]{Comparison of composite spectra of FDF galaxies in the range  
$2.0 < z < 2.5$ differing in $\beta$ and $\gamma_{34}$. Upper spectrum in the 
upper panel: composite of 16 FDF galaxies with $\beta < -0.4$. Lower spectrum 
in the upper panel (offset by $-0.5$ for clarity): composite of eight FDF 
galaxies with $-0.4 < \beta < 0.1$ and $\gamma_{34} < -2$. The lower panel 
gives the ratio of both composites. The positions of prominent UV lines and 
the UV dust feature are indicated.}
\label{fig_demobump}
\end{figure}

\begin{figure}
\centering 
\includegraphics[width=8.8cm,clip=true]{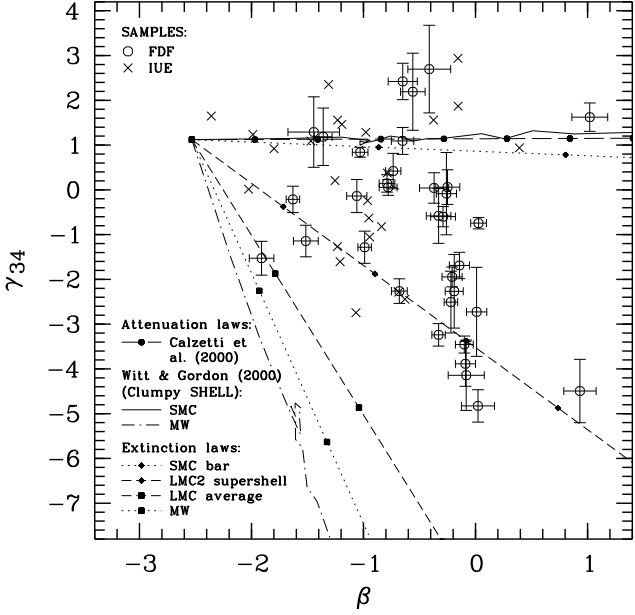}
\caption[]{The UV-bump indicator $\gamma_{34}$ as a function of the 
reddening measure $\beta$ for our local (crosses) and  $2 < z < 2.5$ FDF 
galaxies (circles). The diagram also shows combined stellar population 
synthesis and dust attenuation models as a function of the optical depth of
the dust. The dust attenuation models plotted are the Calzetti et al. 
(\cite{CAL00}) law (long dashed lines and circles) and the \cite{WIT00} 
models for the clumpy SHELL configuration with SMC (solid lines) and MW dust 
(dash-dotted lines). Furthermore, the extinction laws of the SMC bar (dotted 
lines and lozenges), the LMC\,2 supershell (dashed lines and lozenges), the 
LMC average (dashed lines and squares), and the MW (dotted lines and squares) 
are indicated (\cite{CAR89}; Gordon et al. \cite{GOR03}). The symbols are 
plotted in intervals of $\Delta E_{B-V} = 0.1$. In all cases a \cite{MAR05} 
model with Salpeter IMF, continuous star-formation, an age of 200\,Myr, and 
solar metallicity has been taken.}
\label{fig_bumpstrength}
\end{figure}

Almost all the nearby IUE starbursts exhibit $\beta < -0.4$,
the mean value of this parameter being $\langle\beta\rangle = -1.05 \pm 0.13$
(see Table~\ref{tab_gamma_mean}). For this reference sample, 
$\langle\gamma_{34}\rangle = +0.27 \pm 0.32$, i.e. the evidence for the 
presence of the UV bump is weak.

Hence, we divide the $z \sim 2$ UV-luminous FDF galaxies in two subsamples,
with either $\beta < -0.4$ or $-0.4 < \beta < 0.1$. We find that
the FDF subsample of 16 objects with $\beta < -0.4$ exhibits
$\langle\beta\rangle = -1.01 \pm 0.11$
and $\langle\gamma_{34}\rangle = +0.36 \pm 0.36$, in good agreement with
the reference sample at $z \sim 0$. Conversely, the FDF subsample of 16 
objects with $-0.4 < \beta < 0.1$ exhibits 
$\langle\beta\rangle = -0.18 \pm 0.03$ and 
$\langle\gamma_{34}\rangle = -2.04 \pm 0.40$. This confirms that
the carriers of the 2175\,\AA{} dust absorption feature are present
in red (i.e. with $\beta > -0.4$) UV-luminous FDF galaxies at $z \sim 2$.
Nevertheless, objects with MW-type dust seem not to be present among
the FDF sample galaxies as well as in the IUE sample.

The presence of the UV-bump can be appreciated directly from the composite
spectrum of the eight FDF galaxies with $\beta > -0.4$ and $\gamma_{34} < -2$ 
($\langle\gamma_{34}\rangle = -3.4$), reproduced in Fig.~\ref{fig_demobump}.
As a reference, the composite spectrum of the 16 FDF galaxies with 
$\beta < -0.4$ ($\langle\gamma_{34}\rangle = 0.4$) is also shown. The two 
spectra are normalised in the wavelength range from 1250 to 1500\,\AA{}. The 
ratio between the ``blue'' and ``red'' composite spectra exhibits a broad 
feature centred at $2166 \pm 15$\,\AA{}, consistent with the centre of the UV 
bump. Its slope is negative owing to the different values of $\beta$ for the 
two subsamples.

Having established the presence of a UV bump in the extinction curves of the 
red UV-luminous FDF galaxies at $z \sim 2$, can we also estimate its 
strength? Unfortunately, there are no \cite{WIT00} models for extinction 
curves different from those typical either of the diffuse ISM of the Milky 
Way or of the SMC bar. Therefore, for illustrational purposes only, we show 
the distribution in the $\beta - \gamma_{34}$ plane of models where dust 
attenuation is described by a non-scattering homogeneous dust screen for 
different empirical extinction laws as a function of $E_{B-V}$ (see Calzetti 
et al. \cite{CAL00}; Gordon et al. \cite{GOR03}). In particular, we consider 
the extinction functions determined for the SMC bar, the LMC\,2  supershell 
region, the average LMC (Gordon et al. \cite{GOR03}), and the average MW 
(\cite{CAR89}), that are displayed in Fig.~\ref{fig_extinct}. As a reference, 
we take the attenuation functions produced by the two-phase, clumpy SHELL 
models with MW- or SMC-type dust of \cite{WIT00}, and the Calzetti et al. 
(\cite{CAL00}) attenuation law. This comparison holds only to the purpose of 
comparing slopes of synthetic spectra (see Sect.~\ref{radtransfer}). 
Fig.~\ref{fig_bumpstrength} reproduces the loci of the different models and 
the distribution of the FDF and IUE sample galaxies. It is evident that a 
mixture of dust grains with properties ranging from those typical of the SMC 
bar to those typical of the LMC\,2 supershell region can account for the 
distribution of the data in the $\beta - \gamma_{34}$ plane. We recall that 
the LMC\,2 supershell region is located near the extremely vigorous 30\,Dor 
star-formation region (Misselt et al. \cite{MIS99}). Conversely, extinction 
curves typical of more quiescent star-formation activities like those of the 
average LMC and MW seem not to describe the distribution of the data in the 
previous plane. This is consistent with the nature of actively star-forming 
galaxies of the FDF and IUE sample galaxies (see Sect.~\ref{basics}).

Comparing the $\gamma_{12}$ values for the different subsamples defined above 
(see Table~\ref{tab_gamma_mean}), we do not find a significant relation 
between the far-UV slopes of the extinction function as indicated by 
$\gamma_{12}$ and the bump strength. The moderate increase of $\gamma_{12}$ 
with $\beta$ is consistent with the predictions from the \cite{WIT00} models 
for SMC-type dust, for which the mean values are in better agreement than for 
the MW-type models (see Sect.~\ref{curvature}).

\subsection{Presence of the UV bump and strengths of stellar and interstellar 
lines}\label{relations}

\begin{figure}
\centering 
\includegraphics[width=8.8cm,clip=true]{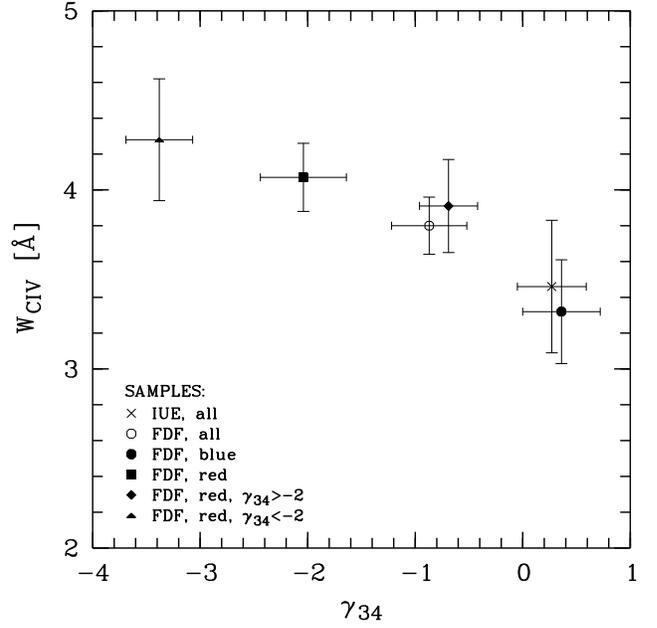}
\caption[]{The EW of the metallicity indicator C\,IV as a function of the 
UV-bump strength indicator $\gamma_{34}$ for our local galaxy sample and 
different subsamples of $2 < z < 2.5$ FDF galaxies (see legend and caption of
Fig.~\ref{fig_gam34bet_mean}). The EWs of C\,IV were derived from composite 
spectra of the different samples. The $1\,\sigma$ error bars include the 
variance between the individual spectra, the errors of the continuum level 
definition, and the statistical uncertainties of the line strengths.}
\label{fig_CIV_gamma34}
\end{figure}

\begin{figure}
\centering 
\includegraphics[width=8.8cm,clip=true]{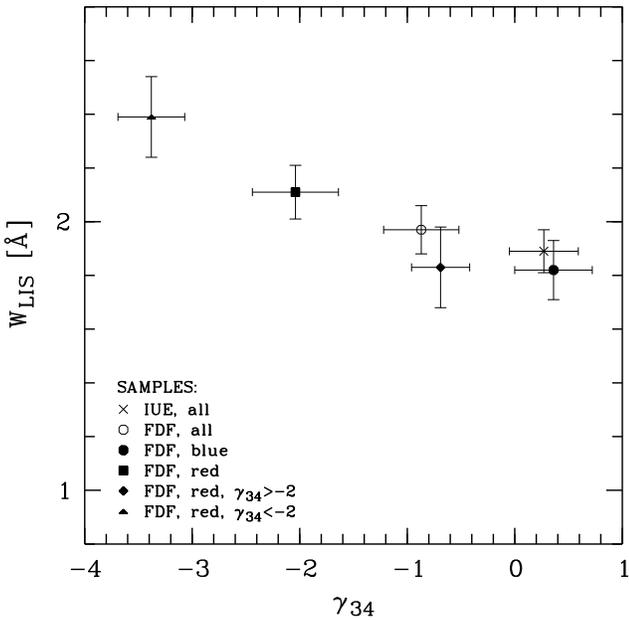}
\caption[]{The average strength of six prominent interstellar low-ionisation 
absorption lines (see Noll et al. \cite{NOL04}) as a function of the UV-bump 
strength indicator $\gamma_{34}$ for our local galaxy sample and different 
subsamples of $2 < z < 2.5$ FDF galaxies (see legend and caption of
Fig.~\ref{fig_gam34bet_mean}). The EWs of the low-ionisation lines were 
derived in the same way as reported in Fig.~\ref{fig_CIV_gamma34} for C\,IV.}
\label{fig_LIS_gamma34}
\end{figure}

Hereafter we investigate whether the presence of the UV bump is linked either 
to the (stellar) metallicity or to properties of the ISM of the system, as 
traced by the equivalent widths of the absorption lines listed in 
Tab.~\ref{tab_ew}. We group the objects in subsamples, as in 
Sect.~\ref{subsamples}.

Fig.~\ref{fig_CIV_gamma34} reproduces $\gamma_{34}$ versus $W_{\rm C\,IV}$.
$W_{\rm C\,IV}$ is a good metallicity indicator in starburst galaxies, the 
C\,IV feature being mainly produced by hot star winds with a strength that 
depends on the chemical composition (Walborn et al. \cite{WAL95}; Heckman et 
al. \cite{HEC98}; Leitherer et al. \cite{LEI01}; Mehlert et al. 
\cite{MEH02}). Interestingly, $W_{\rm C\,IV}$ is smaller for the subsample of 
blue UV-luminous FDF galaxies at $z \sim 2$, as well as for the IUE sample of 
nearby starbursts, than for the red subsample of FDF galaxies. However, red 
FDF galaxies selected according to their values of $\gamma_{34}$ do not 
exhibit significant differences in $W_{\rm C\,IV}$. Hence, the metallicity of 
the OB stars (and, thus, of the ISM where these stars originated) does not 
primarily determine the presence of the carrier of the 2175\,\AA{} dust 
absorption feature among the $z \sim 2$ UV-luminous FDF galaxies. This is 
consistent with studies of extinction curves in the local universe (see 
Sect.~\ref{introduction}). On the other hand, metallicity is higher (by more 
than $2\,\sigma$) for FDF objects affected by a larger reddening at 
rest-frame UV wavelengths (i.e. with a larger $\beta$). This result holds for 
the whole FDF spectroscopic sample (Noll et al. \cite{NOL04}).

Finally, the parameter $\gamma_{34}$ seems to depend also on the average
EW of the six most prominent low-ionisation absorption lines that fall within 
the rest-frame UV domain mapped by FORS ($W_{\rm LIS}$, see 
Fig.~\ref{fig_LIS_gamma34}). In particular, values of $\gamma_{34}$ close to 
zero are associated with low values of $W_{\rm LIS}$. They characterise the 
IUE local starbursts and the blue $z \sim 2$ UV-luminous FDF galaxies. On 
average, the red $z \sim 2$ UV-luminous FDF galaxies exhibit larger values of 
$W_{\rm LIS}$ (by more than $1.5\,\sigma$). This difference is essentially
caused by the red FDF objects with $\gamma_{34} < -2$, having larger values 
of $W_{\rm LIS}$ at almost the $3\,\sigma$ level.

This behaviour can be interpreted as follows. The low-ionisation absorption
lines trace the spatial and velocity distribution of the neutral interstellar
gas illuminated by the rest-frame UV continuum along the observer's line of 
sight. In particular, $W_{\rm LIS}$ is a proxy for the neutral clouds' 
covering fraction (Shapley et al. \cite{SHA03}). Larger values of 
$W_{\rm LIS}$ imply a larger obscuration at rest-frame UV wavelengths, if the 
spatial distributions of neutral gas and dust are correlated. Hence, they 
imply more negative values of $\gamma_{34}$, if the extinction curve exhibits 
a UV bump, as shown by Fig.~\ref{fig_gam34bet_mean} for the dusty media 
modelled by \cite{WIT00}.

\section{Discussion}\label{discussion}

\begin{figure}
\centering 
\includegraphics[width=8.8cm,clip=true]{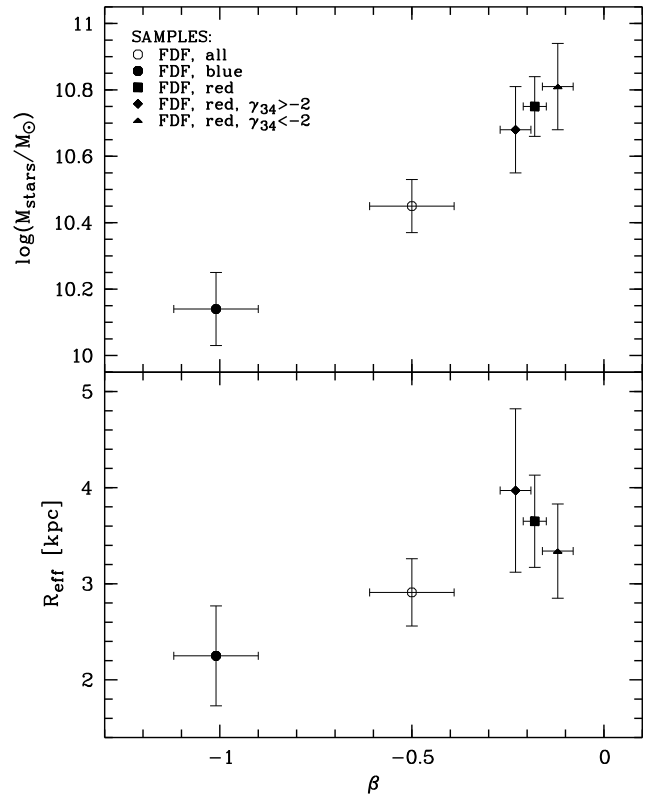}
\caption[]{The logarithm of the total stellar mass (Drory et al. 
\cite{DRO05}) in solar units (upper panel) and the half-light radius 
(Pannella et al. \cite{PAN05}) in kpc (lower panel) as a function of the 
reddening measure $\beta$ for different subsamples of $2 < z < 2.5$ FDF 
galaxies (see legend and caption of Fig.~\ref{fig_gam34bet_mean}). The 
$1\,\sigma$ error bars are based on the variance in mass and radius between 
different galaxies.}
\label{fig_logM_Re_beta}
\end{figure}

Dust grains play a fundamental role in the formation and evolution of 
galaxies. In fact, they are both a byproduct of stellar evolution (in 
particular, a sink of metals in the gas phase), as well as a site for the 
efficient formation of $\rm H_2$ molecules (e.g. Cazaux \& Tielens 
\cite{CAZ02}, \cite{CAZ04}). Furthermore, dust grains absorb stellar light 
and re-emit it at mid-IR--mm wavelengths, controlling the energy balance in 
the ISM and protostellar gas clouds on one hand, and complicating the 
interpretation of observed SEDs of any dusty system on the other. Finally, 
they can offer the surface that is needed for momentum-driven winds on both 
stellar and galactic scales. Those effects of dust already seem to take place
when the metallicity of the environment is very low (i.e. 
$\sim 0.01$\,Z$_{\odot}$, Hirashita \& Ferrara \cite{HIR02}; Morgan \& 
Edmunds \cite{MOR03}). For all those reasons, the first dust enrichment in 
the course of galaxy evolution may drive an enhanced, high star-formation 
rate, shaping the cosmic SFR history at the highest redshifts.

The first source of dust in the universe is Type~II supernovae or pair 
instability supernovae, beginning to occur a few Myr after the onset of 
star-formation. After a few $10^2$\,Myr evolved intermediate/low mass stars 
start to contribute to the dust enrichment. Hence, for the individual galaxy, 
the relative importance of the two sources of dust depends on time and SFR 
history (see, e.g., Dwek \cite{DWE98}).

The extinction curves expected in young (i.e. less than 1\,Gyr old) galaxies
are quite sensitive to internal metal mixing in supernovae (Hirashita et al.
\cite{HIR05}). In general, these extinction curves do not exhibit a 
2175\,\AA{} absorption feature (as the standard MW extinction curve instead)
because the contribution of carbon grains is small (for dust yields, see 
Todini \& Ferrara \cite{TOD01}; Nozawa et al. \cite{NOZ03}; Schneider et al. 
\cite{SCHN04}). Furthermore, they are different from the MW, LMC, and SMC 
extinction curves, especially at rest-frame UV wavelengths (Hirashita et al. 
\cite{HIR05}). This is no surprise, since the ages of the MW, LMC, and SMC 
are much longer than 1\,Gyr and, thus, at least a change in the dust 
production mechanism has taken place. Furthermore, the dusty gas outflows of 
intermediate/low mass stars are carbon rich.

At present, there are only few observational studies on the extinction curves 
in systems at very high redshift. The extinction properties of broad 
absorption line quasars at $z > 4.9$ seem to be different from those of 
analogous objects at $z < 4$ (Maiolino et al. \cite{MAI04a}). In particular, 
the extinction curve of SDSS J\,1048$+$4637 at $z = 6.2$ has been shown to be 
fitted by dust models of Type~II supernovae rather than those of pair 
instability supernovae (Maiolino et al. \cite{MAI04b}). For ages of the 
universe much longer than 1\,Gyr (i.e. for $z << 5$), the known extinction 
curves provide controversial conclusions as for the presence of the 
2175\,\AA{} absorption feature (see Sect.~\ref{introduction}). This may be a 
consequence of at least a difference in the ``characteristic'' SFR history of 
different classes of objects.

In particular, the absence of a pronounced UV bump seems to characterise 
the average extinction curve in the large, well studied photometric 
sample of LBGs at $z \sim 3$ observed by Steidel et al. (\cite{STE03}) (Vijh 
et al. \cite{VIJ03}). Interestingly, the UV bump is weak if not absent also 
in starburst galaxies of the local universe, whatever their metallicity 
(Calzetti et al. \cite{CAL94}; Gordon et al. \cite{GOR97}). Is this so for 
all UV-luminous galaxies down to $z = 0$ then?

In order to answer to this question, we have investigated the extinction 
properties in 34 UV-luminous galaxies at $2 < z < 2.5$ from the FDF 
spectroscopic survey (see Sect.~\ref{FDF}). Using a new parametric 
description of the rest-frame UV spectra (see Sect.~\ref{method}) we have
found that most of the $z \sim 2$ UV-luminous galaxies are characterised by 
an extinction curve that is intermediate between those of the SMC and the MW 
(see Sect.~\ref{results}). This means that the characteristic extinction 
curve of these objects exhibits a weaker 2175\,\AA{} feature but a steeper 
far-UV rise with respect to the average extinction curve for the diffuse ISM 
of the MW.

The bulk of the mixture of dust grains (regarding size and chemical 
composition) present in all the 34 UV-luminous galaxies at $2 < z < 2.5$ is 
different from the mixture of dust grains producing the standard MW 
extinction curve. This is true especially for those galaxies that have 
probably a low dust content (and, thus, a low value of $\beta$ as for our 
models). On the other hand, SMC-like dust may characterise up to 30\% of 
these galaxies. In conclusion, the bulk properties of dust in $z \sim 2$ 
UV-luminous galaxies seem to range between those in the SMC and LMC.

The size of the $z \sim 2$ UV-luminous galaxy sample allows sub-sampling. In 
general, the 18 ``red'' (i.e. with $\beta > -0.4$) UV-luminous galaxies at 
$2 < z < 2.5$ exhibit a stronger evidence of the presence of the UV bump. The 
presence of the UV-bump in the eight UV-luminous galaxies at $z \sim 2$
with $-0.4 < \beta < 0.1$ and $\gamma_{34} < -2$ 
($\langle\gamma_{34}\rangle = -3.4$) can be appreciated directly from the 
ratio of the ``blue'' and ``red'' FDF composite spectra reproduced in 
Fig.~\ref{fig_demobump}. This result is at variance with that obtained for 
the nearby starburst sample, that appears to be characterised by SMC-like 
dust whatever $\beta$, consistent with Gordon et al. (\cite{GOR97}), though 
it mostly exhibits $\beta < -0.4$. Our interpretation of this finding is that 
the carrier of the 2175\,\AA{} feature is more abundant in redder and, thus, 
dustier objects in the $z \sim 2$ UV-luminous galaxy sample.

On the other hand, $z \sim 2$ UV-luminous galaxies with $\beta \sim 0$ 
exhibit a large scatter in both the parameters $\gamma_{12}$ and 
$\gamma_{34}$ (see Fig.~\ref{fig_bumpstrength} and 
Table~\ref{tab_gamma_mean}). Now, $\gamma_{12}$ is particularly sensitive to 
the far-UV slope of the extinction curve, while $\gamma_{34}$ is particularly 
sensitive to the presence (and strength) of the UV bump. We conclude that 
there is no evidence of a correlation between the strength of the far-UV 
slope and the strength of the UV bump, for the extinction curves of the FDF 
galaxies with evidence of the UV bump. This may suggest that different 
environments (in terms of properties of the ISM and of the general radiation 
field) and/or different degrees of dust processing characterise different 
$z \sim 2$ UV-luminous galaxies, if the lessons derived from the local 
universe apply to the universe at $z \sim 2$ (see Sect.~\ref{introduction}).

The available data allow us to check the first suggestion. We find that the 
FDF sample galaxies with $\beta \sim 0$ are characterised by relatively high 
metallicities of $Z \sim 0.5 - 1$\,Z$_{\odot}$ (see Mehlert et al. 
\cite{MEH02}, \cite{MEH05}). However, the metallicity indicator 
$W_{\rm C\,IV}$ does not show any significant correlation with $\gamma_{34}$ 
(see Fig.~\ref{fig_CIV_gamma34}). Hence, a relatively high metallicity (i.e. 
close to the solar value) seems to be a prerequisite for the occurrence of 
the UV bump. However, it does not determine the strength of the 2175\,\AA{} 
feature.

Furthermore, the $z \sim 2$ UV-luminous galaxies with evidence of a 
relatively strong UV bump are among the most massive and largest galaxies at 
this redshift. This is shown in Fig.~\ref{fig_logM_Re_beta}, where we plot 
the estimated total stellar mass\footnote{The total stellar mass was 
estimated by Drory et al. (\cite{DRO05}) following Brinchmann \& Ellis 
(\cite{BRI00}). In particular, Drory et al. (\cite{DRO05}) derived stellar 
masses after applying a large set of two-component models to the whole FDF
photometric data (see Heidt et al. \cite{HEI03}). Each model consisted of a 
combination of two composite stellar populations, as given by a young 
(100\,Myr) burst with a constant SFR plus an older burst ($\ge 500$\,Myr)
with an exponentially declining SFR. Both stellar components were attenuated
independently according to the ``Calzetti law'' (Calzetti et al. 
\cite{CAL00}).} (upper panel) and the rest-frame UV effective 
radius\footnote{Effective radii and ellipticities were estimated by Pannella 
et al. (\cite{PAN05}), who determined the morphology of the whole FDF 
galaxies by applying the publicly available two-dimensional surface 
brightness profile fitting packages GIM2D (Simard et al. \cite{SIM99}) and 
GALFIT (Peng et al. \cite{PEN02}) to publicly available Hubble ACS images of 
the FDF in the F814W broad-band filter. The effective wavelength of this 
filter corresponds to rest-frame wavelengths of about 2500\,\AA{} at 
$z \sim 2$, as for the sample under investigation.} (lower panel) versus 
$\beta$. Nevertheless, there is no correlation between stellar mass or size 
of a galaxy and the strength of the UV bump. In conclusion, an extinction 
curve exhibiting the UV bump is likely to be associated with the most 
massive, largest, and, possibly, more evolved systems among the $z\sim 2$ 
UV-luminous ones.

Finally, we explore the possibility that the topology of the dusty ISM 
produces the large scatter of $\gamma_{34}$ observed for the red UV-luminous 
galaxies at $2 < z < 2.5$. This hypothesis is made appealing by the fact that
the red UV-luminous galaxies with $\gamma_{34} < -2$ exhibit a significantly
larger $W_{\rm LIS}$ than the average $z \sim 2$ UV-luminous galaxy and the 
average local starburst galaxy. We recall that $W_{\rm LIS}$ traces the 
covering fraction of the young stellar population by (dusty) neutral gas 
clouds in the direction to the outside observer. A larger covering fraction 
may be due to inclination effects, for a disk geometry (e.g. Pierini et al. 
\cite{PIE04}). For the $z \sim 2$ FDF objects with $-0.4 < \beta < 0.1$ we 
find small but insignificant differences in the average ellipticities 
$\epsilon$\,\footnote{$\epsilon = \sqrt{1 - b^2 / a^2}$, $a$ = major axis, 
$b$ = minor axis; for the origin of the data see Pannella et al. 
(\cite{PAN05}) and footnote~6} for the $\gamma_{34} > -2$ and 
$\gamma_{34} < -2$ subsamples (see Table~\ref{tab_gamma_mean}), indicating 
$\langle \epsilon \rangle = 0.41 \pm 0.07$ and $0.55 \pm 0.08$, respectively. 
Hence, larger values of $W_{\rm LIS}$ possibly imply a better self-shielding 
of the dust grains that populate the intervening gas clouds, despite 
inclination effects may be not negligible. The carriers of the 2175\,\AA{} 
feature would then be protected from strong and hard radiation fields by a 
screen of other (more robust) dust grains (e.g. Gordon et al. \cite{GOR03}). 
In this case, light emitted by young stars and scattered by the embedding 
dusty gas clouds plus light emitted by stars living in relatively quiescent 
environments (e.g. the diffuse ISM) would dominate the UV spectra. On the 
other hand, smaller values of $W_{\rm LIS}$ may imply a larger dust-clearing 
effect by superwinds (e.g. Strickland \& Stevens \cite{STR00}) in the 
direction to the outside observer. The effect of superwinds is probably 
larger in less evolved systems. In this case, the UV spectra would be 
dominated by light produced in star-formation regions, characterised by 
strong radiation fields and shocks. In these harsh environments, probably the 
carriers of the 2175\,\AA{} feature would have a larger probability to be 
destroyed, if ever produced. However, the local universe shows that the 
latter conclusion is not so straightforward (see Sect.~\ref{introduction}).

\section{Conclusions}\label{conclusions}

We have investigated the properties of the extinction curve in the rest-frame 
UV for a sample of 34 UV-luminous galaxies at redshift $2 < z < 2.5$ selected 
from the FORS Deep Field spectroscopic survey (Noll et al. \cite{NOL04}). Our 
analysis makes use of a new parametric description of the rest-frame UV 
spectra of the observed objects and of synthetic spectra. The latter spectra 
are computed with models combining stellar population evolutionary synthesis 
(Maraston 2005, \cite{MAR05}) and radiative transfer of the stellar and 
scattered radiation through different dusty interstellar media for the SHELL 
dust/stars configuration, in a spherically symmetric geometry (Witt \& Gordon 
2000, \cite{WIT00}).

Firstly, we have established how much the new parameters (seven in total) 
depend on the properties of the stellar populations or of dust attenuation 
assumed for galaxy models aimed at reproducing the rest-frame UV spectra of 
star-forming galaxies.

Secondly, the robustness of our method has been tested against a reference 
sample of 24 nearby starbursts. Comfortably, we reproduce the result that 
local starburst galaxies possibly contain dust with an extinction curve 
lacking the 2175\,\AA{} absorption feature (or ``UV bump''), like the Small 
Magellanic Cloud (SMC) but at variance with the Milky Way (MW), and a steep 
far-UV rise, intermediate between those in the SMC and the MW extinction 
curves (Calzetti et al. \cite{CAL94}; Gordon et al. \cite{GOR97}).

The application of the same method to the sample of $z \sim 2$ UV-luminous
galaxies leads to the following results.
\begin{itemize}
\item The $z \sim 2$ UV-luminous galaxies at $2 < z < 2.5$ with a relatively 
low amount of reddening at rest-frame UV wavelengths (i.e. with a far-UV 
slope $\beta < -0.4$) contain a mixture of dust grains that is similar to 
that held to be present in local starburst galaxies. This analogy holds 
despite the fact that the former galaxies are about one order of magnitude 
more luminous than the latter.
\item For the highly reddened (i.e. with $-0.4 < \beta < 0.1$) $z \sim 2$
UV-luminous galaxies, the extinction curves range between those typical of 
the SMC and Large Magellanic Cloud (LMC). Therefore, part of these objects
exhibits UV bumps as pronounced as in the LMC.
\item The amount of reddening at rest-frame UV wavelengths is larger in more 
metal-rich objects, the equivalent width of the C\,IV doublet 
($W_{\rm C\,IV}$) being used as an indicator of the total metallicity. On the 
other hand, objects with large values of $W_{\rm C\,IV}$ (i.e. metal rich) 
can exhibit both SMC- and LMC-type dust, despite the 2175\,\AA{} absorption 
feature is associated with the presence of carbonaceous dust grains.
\item The presence of the UV bump seems to be associated with a large average
equivalent width of the six most prominent, rest-frame UV, low-ionisation 
absorption lines falling in the FORS spectra. The average equivalent width of 
these saturated lines offers a proxy for the covering fraction of young 
massive stars by neutral gas clouds (containing dust grains), and, thus, for 
the ISM topology.
\item Furthermore, we find that the most opaque and, thus (for our models), 
dustiest UV-luminous FDF galaxies at $z \sim 2$ tend to be among the most 
metal rich (i.e. with an average metallicity 
$\langle Z \rangle \sim 0.5 - 1$\,Z$_{\odot}$), most massive (i.e. with an 
average stellar mass 
$\langle M_{\rm stars} \rangle \sim 6 \times 10^{10}$\,M$_{\odot}$), and 
largest (i.e. with an average rest-frame UV effective radius 
$\langle R_{\rm eff} \rangle \sim 4$\,kpc) systems.
\end{itemize}

We interpret these results as the evidence for a difference in the properties
of the dusty ISM among the most evolved UV-luminous, massive galaxies at 
$z \sim 2$. As for the local universe, the diversity in the bulk properties
of the dust grains from one galaxy to another is possibly caused by the sum
of the local effects due to stellar radiation field, shocks, ISM topology,
and history of chemical enrichment. In particular, the degree of 
self-shielding by dust may play an important role for the production and/or 
survival of the small carbonaceous grains that are held to be the carriers of
the UV bump already at $z \sim 2$.

\begin{acknowledgements}
First of all, we thank Claudia Maraston for calculating special composite 
stellar population synthesis models for our project and for reading the 
manuscript. Moreover, thanks go to Niv Drory and Maurilio Pannella for 
providing mass estimates and morphological information for our FDF sample. We
thank Karl D. Gordon for a stimulating discussion. We are also grateful to 
Immo Appenzeller, Ulrich Hopp, D\"orte Mehlert, and Christian Tapken for 
carefully reading the manuscript and for their valuable comments. Finally, we 
thank the referee for her/his receptive comments. The investigation of our 
sample of local starburst galaxies has been based on INES data from the IUE 
satellite and has made use of the NASA/IPAC Extragalactic Database (NED) 
which is operated by the Jet Propulsion Laboratory, California Institute of 
Technology, under contract with the National Aeronautics and Space 
Administration. This research was supported by the German Science Foundation 
(DFG, SFB 375).  
\end{acknowledgements}

\end{document}